\documentclass[modern]{aastex62}

\usepackage{xcolor}

\submitjournal{PASP}

\shorttitle{CSP-II: Near-Infrared Hubble Diagram for SNe~Ia to $z \sim 0.1$}
\shortauthors{Phillips et al.}

\begin{document}

\title{Carnegie Supernova Project-II: Extending the Near-Infrared Hubble Diagram 
for Type~Ia Supernovae to $z \sim 0.1$\footnote{This paper includes data gathered with the 6.5 meter 
Magellan telescopes at Las Campanas Observatory, Chile.}}

\correspondingauthor{Mark M. Phillips}
\email{mmp@lco.cl}

\author[0000-0003-2734-0796]{M.~M.~Phillips}
\affiliation{Carnegie Observatories, Las Campanas Observatory, Casilla 601, La Serena, Chile}

\author[0000-0001-6293-9062]{Carlos~Contreras}
\affiliation{Carnegie Observatories, Las Campanas Observatory, Casilla 601, La Serena, Chile}

\author[0000-0003-1039-2928]{E.~Y.~Hsiao}
\affiliation{Carnegie Observatories, Las Campanas Observatory, Casilla 601, La Serena, Chile}
\affiliation{Department of Physics and Astronomy, Aarhus University, Ny Munkegade 120, DK-8000 Aarhus C, Denmark}
\affiliation{Department of Physics, Florida State University, 77 Chieftan Way, Tallahassee, FL  32306, USA}

\author[0000-0003-2535-3091]{Nidia~Morrell}
\affiliation{Carnegie Observatories, Las Campanas Observatory, Casilla 601, La Serena, Chile}

\author[0000-0003-4625-6629]{Christopher R.~Burns}
\affiliation{Observatories of the Carnegie Institution for Science, 813 Santa Barbara St., Pasadena, CA 91101, USA}

\author[0000-0002-5571-1833]{Maximilian Stritzinger}
\affiliation{Department of Physics and Astronomy, Aarhus University, Ny Munkegade 120, DK-8000 Aarhus C, Denmark}

\author[0000-0002-5221-7557]{C.~Ashall}
\affiliation{Department of Physics, Florida State University, 77 Chieftan Way, Tallahassee, FL  32306, USA}

\author[0000-0003-3431-9135]{Wendy~L.~Freedman}
\affiliation{Observatories of the Carnegie Institution for Science, 813 Santa Barbara St., Pasadena, CA 91101, USA}
\affiliation{Department of Astronomy and Astrophysics, University of Chicago, 5640 S. Ellis Ave, Chicago, IL 60637, USA}

\author[0000-0002-4338-6586]{P.~Hoeflich}
\affiliation{Department of Physics, Florida State University, 77 Chieftan Way, Tallahassee, FL  32306, USA}

\author[0000-0003-0554-7083]{S.~E.~Persson}
\affiliation{Observatories of the Carnegie Institution for Science, 813 Santa Barbara St., Pasadena, CA 91101, USA}

\author[0000-0001-6806-0673]{Anthony L. Piro}
\affiliation{Observatories of the Carnegie Institution for Science, 813 Santa Barbara St., Pasadena, CA 91101, USA}

\author[0000-0002-8102-181X]{Nicholas~B.~Suntzeff}
\affiliation{George P. and Cynthia Woods Mitchell Institute for Fundamental Physics and Astronomy, Texas A\&M University,
Department of Physics and Astronomy,  College Station, TX 77843, USA}

\author[0000-0002-9413-4186]{Syed~A.~Uddin}
\affiliation{Observatories of the Carnegie Institution for Science, 813 Santa Barbara St., Pasadena, CA 91101, USA}

\author{Jorge~Anais}
\affiliation{Carnegie Observatories, Las Campanas Observatory, Casilla 601, La Serena, Chile}

\author[0000-0001-5393-1608]{E. Baron}
\affiliation{University of Oklahoma 440 W. Brooks, Rm 100, Norman, Oklahoma, 73019, USA}

\author[0000-0001-9952-0652]{Luis~Busta}
\affiliation{Carnegie Observatories, Las Campanas Observatory, Casilla 601, La Serena, Chile}

\author{Abdo~Campillay}
\affiliation{Carnegie Observatories, Las Campanas Observatory, Casilla 601, La Serena, Chile}
\affiliation{Departamento de F\'{i}sica, Universidad de La Serena, Cisternas 1200, La Serena, Chile}

\author{Sergio~Castell\'{o}n}
\affiliation{Carnegie Observatories, Las Campanas Observatory, Casilla 601, La Serena, Chile}

\author{Carlos~Corco}
\affiliation{Carnegie Observatories, Las Campanas Observatory, Casilla 601, La Serena, Chile}
\affiliation{SOAR Telescope, Casilla 603, La Serena, Chile}

\author[0000-0002-0805-1908]{T.~Diamond}
\affiliation{Department of Physics, Florida State University, 77 Chieftan Way, Tallahassee, FL  32306, USA}
\affiliation{Laboratory of Observational Cosmology, Code 665, NASA Goddard Space Flight Center, Greenbelt, MD 20771, USA}

\author[0000-0002-8526-3963]{Christa~Gall}
\affiliation{Department of Physics and Astronomy, Aarhus University, Ny Munkegade 120, DK-8000 Aarhus C, Denmark}
\affiliation{Dark Cosmology Centre, Niels Bohr Institute, University of Copenhagen, Juliane Maries Vej 30, 2100 Copenhagen, Denmark}

\author{Consuelo~Gonzalez}
\affiliation{Carnegie Observatories, Las Campanas Observatory, Casilla 601, La Serena, Chile}

\author{Simon Holmbo}
\affiliation{Department of Physics and Astronomy, Aarhus University, Ny Munkegade 120, DK-8000 Aarhus C, Denmark}

\author[0000-0002-6650-694X]{Kevin~Krisciunas}
\affiliation{George P. and Cynthia Woods Mitchell Institute for Fundamental Physics and Astronomy, Texas A\&M University,
Department of Physics and Astronomy,  College Station, TX 77843, USA}

\author{Miguel~Roth},
\affiliation{Carnegie Observatories, Las Campanas Observatory, Casilla 601, La Serena, Chile}
\affiliation{GMTO Corporation, Presidente Riesco 5335, Of. 501, Nueva Las Condes, Santiago}

\author{Jacqueline~Ser\'{o}n}
\affiliation{Carnegie Observatories, Las Campanas Observatory, Casilla 601, La Serena, Chile}
\affiliation{Cerro Tololo Inter-American Observatory, Casilla 603, La Serena, Chile}

\author[0000-0002-2387-6801]{F.~Taddia}
\affiliation{The Oskar Klein Centre, Department of Astronomy, Stockholm University, SE-106 91 Stockholm, Sweden}

\author{Sim\'{o}n~Torres}
\affiliation{SOAR Telescope, Casilla 603, La Serena, Chile}

\author[0000-0003-0227-3451]{J.~P.~Anderson}
\affiliation{European Southern Observatory, Alonso de C\'{o}rdova 3107, Casilla 19, Santiago, Chile}

\author[0000-0003-0424-8719]{C.~Baltay}
\affiliation{Department of Physics, Yale University, 217 Prospect Street, New Haven, CT 06511, USA}

\author[0000-0001-5247-1486]{Gast\'{o}n Folatelli}
\affiliation{Facultad de Ciencias Astron\'{o}micas y Geof\'{i}sicas, Universidad Nacional de La Plata, Instituto de Astrof\'{i}sica de La Plata (IALP), 
CONICET, Paseo del Bosque S/N, B1900FWA La Plata, Argentina}

\author[0000-0002-1296-6887]{L.~Galbany}
\affiliation{PITT PACC, Department of Physics and Astronomy, University of Pittsburgh, Pittsburgh, PA 15260, USA}

\author[0000-0002-4163-4996]{A.~Goobar}
\affiliation{The Oskar Klein Centre, Department of Physics, Stockholm University, SE-106 91 Stockholm, Sweden}

\author{Ellie~Hadjiyska}
\affiliation{Department of Physics, Yale University, 217 Prospect Street, New Haven, CT 06511, USA}

\author[0000-0001-7981-8320]{Mario~Hamuy}
\affiliation{Universidad de Chile, Departamento de Astronom\'{\i}a, Casilla 36-D, Santiago, Chile} 

\author[0000-0002-5619-4938]{Mansi Kasliwal}
\affiliation{Caltech, 1200 East California Boulevard, MC 249-17, Pasadena, CA 91125, USA}

\author[0000-0003-1731-0497]{C.~Lidman}
\affiliation{The Research School of Astronomy and Astrophysics, Australian National University, ACT 2601, Australia}

\author[0000-0002-3389-0586]{Peter~E.~Nugent}
\affiliation{Lawrence Berkeley National Laboratory, Department of Physics, 1 Cyclotron Road, Berkeley, CA 94720, USA}
\affiliation{Astronomy Department, University of California at Berkeley, Berkeley, CA 94720, USA}

\author[0000-0002-4436-4661]{S.~Perlmutter}
\affiliation{Lawrence Berkeley National Laboratory, Department of Physics, 1 Cyclotron Road, Berkeley, CA 94720, USA}
\affiliation{Astronomy Department, University of California at Berkeley, Berkeley, CA 94720, USA}

\author{David~Rabinowitz}
\affiliation{Department of Physics, Yale University, 217 Prospect Street, New Haven, CT 06511, USA}

\author[0000-0003-4501-8100]{Stuart~D.~Ryder}
\affiliation{Department of Physics \& Astronomy, Macquarie University, NSW 2109, Australia}

\author[0000-0001-6589-1287]{Brian~P.~Schmidt}
\affiliation{The Research School of Astronomy and Astrophysics, Australian National University, ACT 2601, Australia}

\author[0000-0003-4631-1149]{B.~J.~Shappee}
\affiliation{Institute for Astronomy, University of Hawaii, 2680 Woodlawn Drive, Honolulu, HI 96822, USA}

\author{Emma~S.~Walker}
\affiliation{Department of Physics, Yale University, 217 Prospect Street, New Haven, CT 06511, USA}





\begin{abstract}
The Carnegie Supernova Project-II (CSP-II) was an NSF-funded, four-year program to obtain 
optical and near-infrared observations of a ``Cosmology'' sample of $\sim100$ Type~Ia supernovae 
located in the smooth Hubble flow ($0.03 \la z \la 0.10$). Light curves were also obtained of a 
``Physics'' sample composed of 90 nearby Type~Ia supernovae at $z \leq 0.04$ 
selected for near-infrared spectroscopic time-series observations.  The primary emphasis of the 
CSP-II is to use the combination of optical and near-infrared photometry to achieve a 
distance precision of better than 5\%.  In this 
paper, details of the supernova sample, the observational strategy, and the characteristics of the 
photometric data are provided.  In a companion paper, the near-infrared spectroscopy 
component of the project is presented.
\end{abstract}




\section{Introduction}
\label{sec:intro}

A key goal of observational cosmology is to constrain the nature of dark energy 
through the detailed, accurate, and unbiased measurement of the  expansion history 
of the Universe. Einstein's cosmological constant, for which the dark energy equation of 
state parameter, $w$, is precisely $-1$, is entirely consistent with the most recent results
\citep{betoule14,scolnic18}.  However, so are several competing models that are as
fundamentally different from each other as they are to the cosmological constant
\citep[e.g., see the review by][]{yoo12}.
Currently, the tightest limits on the value of $w$ come from combining observations of Type~Ia 
supernovae (SNe~Ia) with data from other probes, such as the Cosmic Microwave 
Background and galaxy clustering. Excluding SNe~Ia from these analyses results in 
considerably weaker constraints on $w$ \citep[e.g.,][]{sullivan11,betoule14}.  Nevertheless,
the power of experiments such as the Baryon Oscillation Spectroscopic Survey 
\citep[BOSS;][]{dawson13} and its successor, the Dark Energy Spectroscopic Instrument
\citep[DESI;][]{desi16}, to constrain $w$ is expected to soon match and perhaps exceed the 
power of the current state-of-the-art SNe~Ia experiments.

Improving the experiments with SNe~Ia 
is not just a question of observing more SNe~Ia since any survey, 
no matter how large, will ultimately be limited by systematic errors related to both the 
photometric calibration and the physical nature of SNe~Ia \citep[e.g.,][]{conley11,scolnic18}.
In the optical, SNe Ia are not perfect standard candles. Rather, their successful use in 
cosmology is due to the discovery of empirical relations between luminosity, light-curve
decline rate, and color that dramatically decrease the  dispersion in peak luminosities
\citep{pskovskii77,phillips93,tripp98}. These relationships reduce the intrinsic Hubble diagram 
scatter typically to $\sim$0.15~mag ($\sim$7\% in distance). A luminosity correction 
dependent on host-galaxy mass has also been introduced \citep{kelly10,lampeitl10,sullivan10}. 
Recently, this effect has been confirmed by \citet{uddin17} using a sample of 1338 SNe~Ia.
Interestingly, these authors found that SNe Ia in hosts with high specific star formation rates 
display the lowest intrinsic dispersion ($0.08 \pm 0.01$~mag) in luminosity after correction 
for light-curve decline rate and host galaxy mass \citep[see also][]{rigault13}.

Reducing the Hubble diagram scatter even further is highly desirable, as it directly leads 
to tighter cosmological constraints. However, at optical wavelengths, reducing the scatter has 
proven difficult, despite many years of effort. Fortunately, observations in the near-infrared (NIR) 
offer a way forward. This is because extinction from dust is reduced in the NIR and because 
SNe~Ia in the NIR are intrinsically better standard candles \citep{elias85,meikle00}.  NIR 
observations may also avoid possible dimming by dust in the intergalactic medium, currently 
only poorly constrained with optical observations \citep{goobar18}. The potential of SNe Ia as 
distance indicators in the NIR has been clearly demonstrated by \citet{krisciunas04b}, 
\citet{wood-vasey08}, \citet{mandel11}, \citet{barone-nugent12}, and \citep{stanishev18}, and 
by the extensive observations of the first phase of the Carnegie Supernova Project 
\citep[CSP-I;][]{folatelli10,kattner12,phillips12,burns14,burns18}.

The CSP-I was an NSF-funded project initiated in September 2004 to establish a fundamental 
data set of optical and NIR light curves of SNe~Ia in a well-defined and understood photometric 
system \citep{hamuy06}.  The CSP-I optical imaging was obtained on over 1000 nights with the 
Las Campanas Observatory (LCO) Henrietta Swope 1~m telescope in the Sloan 
Digital Sky Survey $ugri$ filters and Johnson $BV$ filters. The NIR imaging was obtained with the
Swope telescope and the LCO 2.5~m du Pont telescope, mostly in the $YJH$ bandpasses. 
Over the 5-year duration of the project, optical light curves were obtained for 123 SNe~Ia, 
83 SNe~II, and 34~SNe Ib/Ic/IIb, with NIR photometry having been obtained for $\sim$85\% of these. 
In addition, over 250 SNe of all types (including 129 SNe Ia) were monitored via optical spectroscopy.  
We have found that the data set for the Type~Ia events has allowed us to improve dust extinction 
corrections \citep{burns14,burns18} and to investigate systematic effects in absolute magnitudes 
possibly due to differences in either age or metallicity, or both. The CSP-I observations are also 
being used to gain a deeper understanding of the physics of SNe~Ia 
\citep[e.g.,][]{hoeflich17,hoeflich10,gall18}. 

The CSP-I SN~Ia optical and NIR light curves were published in three data release papers 
\citep{contreras10,stritzinger11b,krisciunas17}, and most of the optical spectra have also been 
published \citep{folatelli13}.  In a recent paper by \citet{burns18}, we have presented a full analysis of 
the $uBgVriYJH$ Hubble diagrams for the CSP-I sample using new intrinsic color relations as 
a function of the $s_{BV}$ color-stretch parameter \citep{burns14}.  
Excluding the $u$ band, which is affected by \ion{Ca}{2} H~\&~K
absorption features \citep{burns14}, we found peculiar-velocity-corrected dispersions of
5--7\% in distance for the full sample of 120 SNe, and 4--6\% for the subset of events with
$s_{BV} > 0.5$ and $(B-V) < 0.5$~mag. However, the median sample redshift of the CSP-I SNe~Ia is 
$z = 0.024$, where the root mean square (rms) effect due to peculiar velocities is $\sim$4\% in 
distance, and therefore is comparable to the intrinsic dispersion that we are attempting to measure. 
This limitation can be overcome by extending observations further into the smooth Hubble flow 
as shown by \citet{barone-nugent12}, who found a dispersion of 0.08~mag (a distance error of 
4\%) for a sample of a dozen SNe~Ia that had redshifts $0.03 < z < 0.09$.  A similar result
was obtained more recently by \citet{stanishev18} for a sample of 16 SNe~Ia in the redshift
range $z = 0.037 - 0.183$. Improving NIR K-corrections offers an 
additional refinement in precision since poorly understood K-corrections directly impact 
the peak magnitudes of the SNe and inflate both statistical and systematic errors \citep{boldt14}.

In 2011, we began a second phase of the Carnegie Supernova Project (CSP-II) to obtain 
optical and NIR observations of SNe~Ia in the smooth Hubble flow. Over a four-year period, 
light curves were obtained for 214 SNe Ia at redshifts $0.004 < z < 0.137$. NIR spectra were 
also obtained for 157 SNe~Ia. This unique data set should provide an essential test of the ultimate 
precision of SNe Ia as distance indicators for determining the local value of the Hubble constant 
and constraining the nature of dark energy. The NIR spectra will also be a valuable resource 
for studying the physics and progenitors of SNe Ia \citep[e.g., see][]{wheeler98}.

In this paper, details of the photometric portion of the CSP-II are presented.  In \S\ref{sec:sample}, 
the CSP-II ``Cosmology'' and ``Physics'' SNe~Ia subsamples are described in more detail,
along with a third homogenous subsample discovered by the La Silla-QUEST (LSQ) supernova 
survey.  Next, in \S\ref{sec:strategy}, the observing strategy is presented.  This is followed in 
\S\ref{sec:opt_phot} and \S\ref{sec:nir_phot} with details of the photometric reductions.  In 
\S\ref{sec:results}, sample optical and NIR light curves representative of the full data set are 
given, followed by a summary of conclusions in \S\ref{sec:conclusions}.

\section{Supernova Subsamples}
\label{sec:sample}

The CSP-II covered four observing seasons spanning 7-8 months each and centered on the 
Chilean summer, beginning in October 2011 and ending in May 2015.  A final total of 214 SNe~Ia 
was observed with redshifts in the range $0.004 < z < 0.137$. Table~\ref{tab:allsne} provides 
basic information for each SN including coordinates, host-galaxy identifications, and heliocentric 
redshifts.  Also listed are the discovery programs from which the SNe were drawn and the 
sources of the classification spectra that identified these events as SNe~Ia.  In cases of multiple
discoveries of the same SN, the name of the survey that first posted the discovery 
is listed first, and the SN is identified by the name given by the first discoverer.  The horizontal 
lines in the table separate the SNe into the main survey groups (ASASSN, CRTS, KISS, LSQ, etc.).  

\begin{figure}[h]
\epsscale{1.0}
\plotone{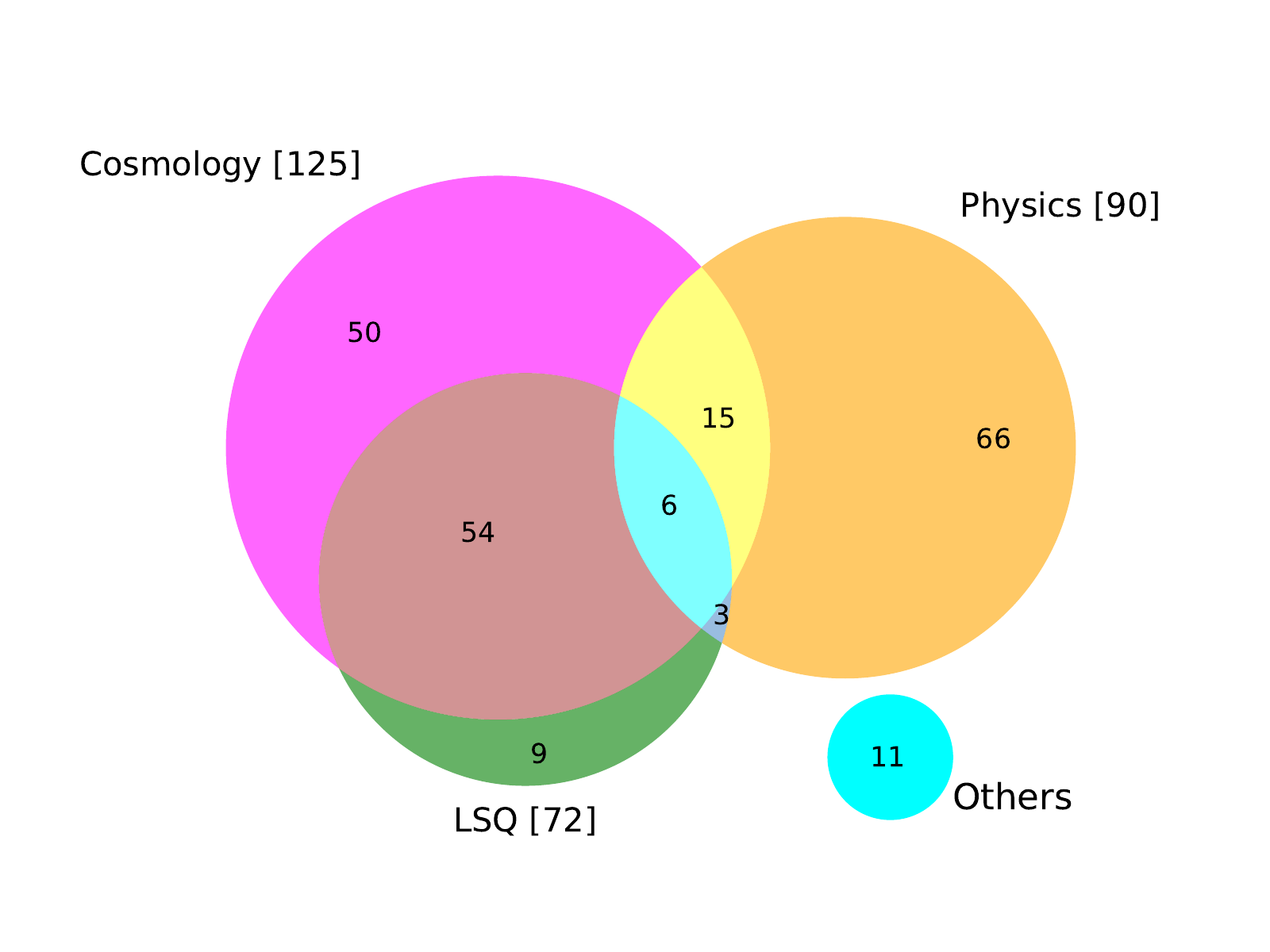} 
\caption{Venn diagram illustrating the subsamples into which the 214 SNe~Ia observed by the
CSP-II are divided.  Note the significant overlap between the Cosmology, Physics, and LSQ
subsamples.}
\label{fig:venn}
\end{figure}

We divide the full sample of 214 SNe~Ia into three subsamples as illustrated in Figure~\ref{fig:venn}
and described in the remainder of this section.

\begin{longrotatetable}
\begin{deluxetable*}{lccllccc}
\tabletypesize{\scriptsize}
\tablecolumns{8}
\tablewidth{0pt}
\tablecaption{CSP-II SNe~Ia\label{tab:allsne}}
\tablehead{
\colhead{SN Name} &
\colhead{R.A. (2000)} &
\colhead{Dec. (2000)} &
\colhead{Host Galaxy} &
\colhead{$z_{\rm{helio}}$\tablenotemark{a}} &
\colhead{Discovery\tablenotemark{b}} &
\colhead{Classification\tablenotemark{b}} &
\colhead{Subsample\tablenotemark{c}}
}
\startdata
ASASSN-14ad                     &   12:40:11.10   &   +18:03:32.80   &   KUG 1237+183                     &   $0.0264$                    &   ASASSN   &  (14)       &   P     \\
ASASSN-14hp                     &   21:30:31.42   &   -70:38:34.35   &   2MASX J21303015-7038489          &   $0.0389$                    &   ASASSN   &   CSP      &   C     \\
ASASSN-14hr                     &   01:50:41.27   &   -14:31:06.37   &   2MASX J01504127-1431032          &   $0.0336$                    &   ASASSN   &   CSP      &   C   \\
ASASSN-14hu                     &   06:43:26.92   &   -69:38:14.70   &   ESO 058- G 012                   &   $0.0216$\tablenotemark{d}   &   ASASSN   &   CSP      &   P     \\
ASASSN-14jc                     &   07:35:35.29   &   -62:46:12.64   &   2MASX J07353554-6246099          &   $0.0113$                    &   ASASSN   &   PESSTO   &   P     \\
ASASSN-14jg                     &   23:33:13.90   &   -60:34:11.50   &   2MASX J23331223-6034201          &   $0.0148$                    &   ASASSN   &   LCOGT    &   P     \\
ASASSN-14jz                     &   18:44:44.34   &   -52:48:05.48   &   GALEXASC J184443.33-524819.2     &   $0.0158$\tablenotemark{d}   &   ASASSN   &   PESSTO   &        \\
ASASSN-14kd                     &   22:53:24.95   &   +04:47:57.30   &   2MASX J22532475+0447583          &   $0.0242$\tablenotemark{d}    &   ASASSN   &   PESSTO   &        \\
ASASSN-14kq                     &   23:45:15.51   &   -29:47:01.14   &   2MASX J23451480-2947009          &   $0.0336$                    &   ASASSN   &   CSP      &   C   \\
ASASSN-14lo                     &   11:51:53.11   &   +18:32:29.00   &   UGC 06837                        &   $0.0199$                    &   ASASSN   &   LCOGT    &   P     \\
ASASSN-14lp                     &   12:45:09.10   &   -00:27:32.49   &   NGC 4666                         &   $0.0051$                    &   ASASSN   &   ASASSN   &   P     \\
ASASSN-14lq                     &   22:57:19.41   &   -20:58:00.76   &   2MASX J22571481-2058014          &   $0.0262$                    &   ASASSN   &  (15)       &   P     \\
ASASSN-14lt                     &   03:11:02.54   &   -13:06:38.76   &   IC 0299                          &   $0.0320$                    &   ASASSN   &   Asiago   &   C   \\
ASASSN-14lw                     &   01:06:49.17   &   -46:58:59.96   &   GALEXASC J010647.95-465904.1     &   $0.0209$\tablenotemark{d}   &   ASASSN   &   CSP      &   P     \\
ASASSN-14me                     &   01:26:40.08   &   -57:59:49.31   &   ESO 113- G 047                   &   $0.0178$\tablenotemark{d}    &   ASASSN   &   ASASSN   &   P     \\
ASASSN-14mf                     &   00:04:54.46   &   -32:26:14.63   &   GALEXASC J000454.54-322615.3     &   $0.0311$                    &   ASASSN   &   ASASSN   &   C   \\
ASASSN-14mw                     &   01:41:25.16   &   -65:37:01.26   &   AM 0139-655 NED02                &   $0.0274$                  &   ASASSN,OGLE   &   ASASSN   &   C,P   \\
ASASSN-14my                     &   11:38:29.98   &   -08:58:35.79   &   NGC 3774                         &   $0.0205$                &   ASASSN,PS1   &   CSP      &   P     \\
ASASSN-15aj                     &   10:52:53.26   &   -32:55:34.86   &   NGC 3449                         &   $0.0109$                    &   ASASSN   &   CSP      &   P     \\
ASASSN-15al                     &   04:57:49.63   &   -21:35:34.11   &   GALEXASC J045749.46-213526.3     &   $0.0338$\tablenotemark{d}   &   ASASSN,Gaia   &   PESSTO   &   C     \\
ASASSN-15as                     &   09:39:16.55   &   +06:25:48.53   &   SDSS J093916.69+062551.1         &   $0.0286$\tablenotemark{d}   &   ASASSN   &   CSP      &   C,P   \\
ASASSN-15ba                     &   14:04:55.09   &   +08:55:14.53   &   SDSS J140455.12+085514.0         &   $0.0231$                    &   ASASSN,CRTS   &   CfA      &   P     \\
ASASSN-15be                     &   02:52:46.39   &   -34:18:52.52   &   GALEXASC J025245.83-341850.6     &   $0.0219$                    &   ASASSN   &   ASASSN   &   P     \\
ASASSN-15bm                     &   15:05:51.58   &   -05:37:37.05   &   LCRS B150313.2-052600            &   $0.0208$\tablenotemark{d}   &   ASASSN   &   ASASSN   &   P     \\
ASASSN-15cb                     &   12:39:50.23   &   +03:47:49.77   &   VCC 1810                         &   $0.0400$                    &   ASASSN,PS1   &   PESSTO   &   C     \\
ASASSN-15cd                     &   09 59 14.75   &   +12 59 20.53   &   CGCG 064-017                     &   $0.0344$                    &   ASASSN   &   PESSTO   &   C     \\
ASASSN-15da                     &   05:23:51.88   &   -24:42:08.38   &   2MASX J05235106-2442201          &   $0.055$\tablenotemark{e}    &   ASASSN   &   ASASSN   &   C     \\
ASASSN-15db                     &   15:46:58.69   &   +17:53:02.55   &   NGC 5996                         &   $0.0110$                    &   ASASSN   &  (16)       &        \\
ASASSN-15dd                     &   15:43:59.07   &   +19:12:40.74   &   CGCG 107-031                     &   $0.0244$                    &   ASASSN   &   ASASSN   &        \\
ASASSN-15eb                     &   08:06:07.40   &   -22:33:48.86   &   ESO 561- G 012                   &   $0.0165$                    &   ASASSN   &   SMT      &   P     \\
ASASSN-15fr                     &   09:20:20.44   &   -07:38:26.78   &   2MASX J09202045-0738229          &   $0.0334$                    &   ASASSN   &   Asiago   &   C,P   \\
ASASSN-15ga                     &   12:59:27.29   &   +14:10:15.79   &   NGC 4866                         &   $0.0066$                    &   ASASSN   &  (17)       &   P     \\
ASASSN-15go                     &   06:11:30.50   &   -16:29:03.52   &   2MASX J06113048-1629085          &   $0.0189$                    &   ASASSN   &   SMT      &   P     \\
ASASSN-15gr                     &   06:45:20.58   &   -34:53:38.11   &   ESO 366- G 015                   &   $0.0243$\tablenotemark{d}   &   ASASSN   &   PESSTO   &   P     \\
ASASSN-15hf                     &   10:29:31.00   &   -35:15:35.60   &   ESO 375- G 041                   &   $0.0062$                    &   ASASSN   &   PESSTO   &   P     \\
ASASSN-15hg                     &   09:53:48.62   &   +09:11:37.78   &   CGCG 063-098                     &   $0.0298$                    &   ASASSN   &  (18)       &   C   \\
ASASSN-15hx                     &   13:43:16.69   &   -31:33:21.55   &   GALEXASC J134316.80-313318.2     &   $0.0083$\tablenotemark{d}    &   ASASSN   &   PESSTO   &   P     \\
PSN J13471211-2422171           &   13:47:12.11   &   -24:22:17.10   &   ESO 509- G 108                   &   $0.0199$                    &   BOSS,ASASSN           &   CSP      &   P     \\
SN2014ao                        &   08:34:33.32   &   -02:32:36.10   &   NGC 2615                         &   $0.0141$                    &   LOSS,ASASSN   &   Asiago   &   P     \\
SN2014I                         &   05:42:19.80   &   -25:32:39.90   &   ESO 487-G36                      &   $0.0300$                    &   BOSS,ASASSN   &   SMT      &   C,P   \\
SN2014eg                        &   02:45:09.27   &   -55:44:16.90   &   ESO 154- G 010                   &   $0.0186$                    &  (1),ASASSN   &   PESSTO   &   P     \\
\hline
CSS111231:145323+025743 (SN2011jt)  &   14:53:23.01   &   +02:57:43.10   &   CGCG 048-051                     &   $0.0278$                    &   CRTS     &   LOSS     &   P     \\
CSS120224:145405+220541 (SN2012aq)  &   14:54:05.13   &   +22:05:40.60   &   SDSS J145405.13+220540.7         &   $0.052$\tablenotemark{e}    &   CRTS     &   Asiago   &   C     \\
CSS120301:162036-102738 (SN2012ar)  &   16:20:36.02   &   -10:27:38.30   &   2MASX J16203650-1028061          &   $0.0283$                    &   CRTS     &   Asiago   &   C,P \\
CSS120325:123816-150632         &   12:38:16.19   &   -15:06:32.15   &   anonymous                        &   $0.0972$\tablenotemark{d}   &   CRTS     &   CRTS     &   C   \\
CSS121114:090202+101800         &   09:02:02.34   &   +10:18:00.10   &   SDSS J090202.18+101759.7         &   $0.0371$                    &   CRTS     &  (19)       &   C     \\
CSS130215:033841+101827 (SN2013ad) &   03:38:41.04   &   +10:18:27.22   &   anonymous                        &   $0.0363$\tablenotemark{d}   &   CRTS     &   CSP      &   C     \\
CSS130303:105206-133424         &   10:52:06.06   &   -13:34:24.70   &   GALEXASC J105206.27-133420.2     &   $0.0789$\tablenotemark{d}   &   CRTS     &   PESSTO   &   C     \\
CSS130315:115252-185920 (SN2013as)  &   11:52:52.34   &   -18:59:19.90   &   anonymous                        &   $0.0685$\tablenotemark{d}   &   CRTS     &   PESSTO   &   C     \\
CSS131031:095508+064831         &   09:55:08.22   &   +06:48:31.40   &   SDSS J095510.00+064830.3         &   $0.0777$                    &   CRTS     &   PESSTO   &   C     \\
CSS140126:120307-010132         &   12:03:06.88   &   -01:01:31.70   &   SDSS J120306.76-010132.4         &   $0.0772$\tablenotemark{d}   &   CRTS     &   PESSTO   &   C     \\
CSS140218:095739+123318         &   09:57:39.11   &   +12:33:17.70   &   SDSS J095738.31+123308.5         &   $0.0773$\tablenotemark{d}   &   CRTS     &   PESSTO   &   C     \\
CSS140914:010107-101840         &   01:01:07.04   &   -10:18:39.90   &   anonymous                        &   $0.03$\tablenotemark{e}     &   CRTS     &   PESSTO   &  C     \\
CSS140925:162946+083831 (SN2014dl)  &   16:29:46.09   &   +08:38:30.60   &   UGC 10414                        &   $0.0330$                    &   CRTS     &   CSP      &   C   \\
CSS150214:140955+173155 (SN2015bo) &   14:09:55.13   &   +17:31:55.60   &   NGC 5490                         &   $0.0162$                    &   CRTS     &   PESSTO   &   P     \\
SNhunt161 (SN2012hl)    &   00:50:17.76   &   +24:31:52.20   &   CSS J005017.69+243154.4          &   $0.0332$\tablenotemark{d}    &   CRTS     &  (20)       &        \\
SNhunt177 (SN2013az)   &   05:39:52.13   &   -40:30:28.10   &   ESO 306-016                      &   $0.0373$                    &   CRTS     &   PESSTO   &   C     \\
SNhunt178 (SN2013bc)  &   13:10:21.31   &   -07:10:24.10   &   IC 4209                          &   $0.0225$                    &   CRTS     &   CSP      &   P     \\
SNhunt188 (SN2013bz)   &   13:26:51.32   &   -10:01:32.20   &   2MASX J13265081-1001263          &   $0.0192$                    &   CRTS     &  (21)       &   P     \\
SNhunt229 (SN2014D)    &   12:10:36.76   &   +18:49:35.40   &   UGC 07170                        &   $0.0082$                    &   CRTS,PS1     &   CSP      &   P     \\
SNhunt281 (SN2015bp)  &   15:05:30.07   &   +01:38:02.40   &   NGC 5839                         &   $0.0041$                    &   CRTS     &  (22)       &   P     \\
SSS111226:125715-172401 (SN2011jn)   &   12:57:14.79   &   -17:24:00.50   &   2MASX J12571157-1724344          &   $0.0475$                    &   CRTS     &  (23)       &   C     \\
MASTER OT J093953.18+165516.4 &   09:39:53.18   &   +16:55:16.40   &   CGCG 092-024                     &   $0.0478$                    &   MASTER,CRTS   &   PESSTO   &   C     \\
SN2014du  &   02:26:23.31   &   +27:39:34.80   &   UGC 01899                        &   $0.0325$                    &   ISSP,CRTS     &   Asiago   &   C,P   \\
\hline
KISS13j (SN2013Y)               &   12:09:39.70   &   +16:12:14.30   &   SDSS J120939.62+161212.2         &   $0.0766$                    &   KISS     &   KISS     &   C     \\
KISS13l (SN2013al)              &   11:14:54.07   &   +29:35:06.00   &   SDSS J111454.06+293508.7         &   $0.1321$                    &   KISS     &  (24)       &   C     \\
KISS13v (SN2013ba)              &   13:52:56.63   &   +21:56:21.70   &   SDSS J135256.58+215621.1         &   $0.080$\tablenotemark{e}    &   KISS     &   KISS     &   C     \\
KISS15m                          &   12:06:00.83   &   +20:36:18.40   &   NGC 4098                         &   $0.0243$                    &   KISS,CRTS     &   CSP      &        \\
SN2015M               &   13:00:32.33   &   +27:58:41.00   &   GALEXMSC J130032.33+275842.3 ?   &   $0.0231$\tablenotemark{g} &   KISS,CRTS,PS1     &   CSP      &   P   \\
\hline
LSQ11bk                         &   04:20:44.25   &   -08:35:55.75   &   anonymous                        &   $0.0403$\tablenotemark{d}   &   LSQ      &   CSP      &   C,L   \\
LSQ11ot                         &   05:15:48.34   &   +06:46:39.36   &   CGCG 421-013                     &   $0.0273$                    &   LSQ      &  (25)       &   C,P,L \\
LSQ11pn (SN2011jq)            &   05:16:41.54   &   +06:29:29.40   &   2MASX J05164149+0629376          &   $0.0327$                    &   LSQ      &   SNF      &   C,P,L \\
LSQ12ca                         &   05:31:03.62   &   -19:47:59.28   &   2MASX J05310364-1948063          &   $0.0994$\tablenotemark{d}   &   LSQ      &   LCOGT    &   L     \\
LSQ12agq                        &   10:17:41.67   &   -07:24:54.45   &   GALEXASC J101741.80-072452.2     &   $0.0642$\tablenotemark{d}   &   LSQ      &   CSP      &   C,L   \\
LSQ12aor                        &   10:55:17.64   &   -14:18:01.38   &   GALEXASC J105517.85-141757.2     &   $0.0934$\tablenotemark{d}   &   LSQ      &   CSP      &   C,L   \\
LSQ12bld                        &   13:42:44.03   &   +08:05:33.74   &   SDSS J134244.72+080531.7         &   $0.0837$                    &   LSQ      &   CSP      &   C,L   \\
LSQ12blp                        &   13:36:05.59   &   -11:37:16.87   &   LCRS B133326.3-112212            &   $0.0743$                    &   LSQ,CRTS      &   CSP      &   C,L   \\
LSQ12btn                        &   09:21:30.47   &   -09:41:29.86   &   2MASX J09213114-0941331          &   $0.0542$                    &   LSQ      &   PESSTO   &   C,L   \\
LSQ12cda                        &   13:50:02.32   &   +09:37:47.10   &   SDSS J135002.40+093755.1         &   $0.1376$                    &   LSQ      &   PESSTO   &   C,L   \\
LSQ12cdl                        &   12:53:39.96   &   -18:30:26.16   &   GALEXASC J125339.85-183025.6     &   $0.1081$\tablenotemark{d}   &   LSQ      &   PESSTO   &   C,L   \\
LSQ12fuk                        &   04:58:15.88   &   -16:17:58.03   &   GALEXASC J045815.88-161800.7     &   $0.0206$\tablenotemark{d}   &   LSQ,CRTS      &   SNF      &   P,L   \\
LSQ12fvl                        &   05:00:50.04   &   -38:39:11.51   &   MCG -06-12-002                   &   $0.0560$                    &   LSQ      &   PESSTO   &   C,L   \\
LSQ12fxd                        &   05:22:17.02   &   -25:35:47.01   &   ESO 487- G 004                   &   $0.0312$                    &   LSQ,CRTS      &   PESSTO   &   C,P,L \\
LSQ12gdj                        &   23:54:43.32   &   -25:40:34.09   &   ESO 472- G 007                   &   $0.0303$                    &   LSQ      &   SNF      &   C,P,L \\
LSQ12gef                        &   01:40:33.70   &   +18:30:36.38   &   2MASX J01403375+1830406          &   $0.0642$\tablenotemark{d}   &   LSQ      &   SNF      &   L     \\
LSQ12gln                        &   05:22:59.41   &   -33:27:51.32   &   GALEXASC J052259.58-332755.3     &   $0.1021$\tablenotemark{d}   &   LSQ      &   PESSTO   &   C,L   \\
LSQ12gpw                        &   03:12:58.24   &   -11:42:40.13   &   2MASX J03125885-1142402          &   $0.0506$\tablenotemark{d}   &   LSQ      &   PESSTO   &   L     \\ 
LSQ12gxj                        &   02:52:57.38   &   +01:36:24.25   &   2MASX J02525699+0136231          &   $0.0355$\tablenotemark{f}   &   LSQ      &   PESSTO   &   L     \\
LSQ12gyc                        &   02:45:50.07   &   -17:55:45.74   &   anonymous                        &   $0.0932$\tablenotemark{f}   &   LSQ      &   PESSTO   &   L     \\ 
LSQ12gzm                        &   02:40:43.61   &   -34:44:25.87   &   GALEXASC J024043.58-344425.0     &   $0.1001$\tablenotemark{d}   &   LSQ      &   PESSTO   &   L     \\
LSQ12hjm                        &   03:10:28.72   &   -16:29:37.08   &   2MASX J03102844-1629333          &   $0.0714$\tablenotemark{d}   &   LSQ      &   SNF      &   L     \\
LSQ12hno                        &   03:42:43.25   &   -02:40:09.76   &   GALEXASC J034243.43-024007.7     &   $0.0473$\tablenotemark{d}   &   LSQ      &   CSP      &   C,L   \\
LSQ12hnr                        &   10:43:14.77   &   -08:46:40.89   &   anonymous                        &   $0.135$\tablenotemark{f}    &   LSQ      &   PESSTO   &   C,L   \\
LSQ12hvj                        &   11:07:38.62   &   -29:42:40.96   &   GALEXASC J110738.65-294235.5     &   $0.0713$\tablenotemark{d}   &   LSQ      &   CSP      &   C,L   \\
LSQ12hxx                        &   03:19:44.23   &   -27:00:25.68   &   2MASX J03194423-2700201          &   $0.0694$                    &   LSQ      &   SNF      &   C,L   \\
LSQ12hzj                        &   09:59:12.43   &   -09:00:08.25   &   2MASX J09591230-0900095          &   $0.0334$\tablenotemark{d}   &   LSQ      &   SNF      &   C,P,L \\
LSQ12hzs                        &   04:01:53.21   &   -26:39:50.15   &   2MASXi J0401529-263947           &   $0.0721$                    &   LSQ      &   CSP      &   C,L   \\
LSQ13lq                         &   13:44:10.81   &   +03:03:43.42   &   SDSS J134410.77+030345.3         &   $0.0757$\tablenotemark{d}   &   LSQ      &   PESSTO   &   C,L   \\
LSQ13pf                         &   13:48:14.35   &   -11:38:38.58   &   LCRS B134534.3-112338            &   $0.0861$\tablenotemark{d}   &   LSQ      &   CSP      &   C,L   \\
LSQ13ry                         &   10:32:48.00   &   +04:11:51.75   &   SDSS J103247.83+041145.5         &   $0.0299$                    &   LSQ      &   CSP      &   C,P,L \\
LSQ13vy                         &   16:06:55.85   &   +03:00:15.23   &   2MASX J16065563+0300046          &   $0.0418$\tablenotemark{d}   &   LSQ      &   LCOGT    &   C,L   \\
LSQ13abo                        &   14:59:21.20   &   -17:09:09.34   &   2MASX J14592124-1709138          &   $0.0675$                    &   LSQ      &   SNF      &   C,L   \\
LSQ13aiz (SN2013cs)             &   13:15:14.81   &   -17:57:55.65   &   ESO 576- G 017                   &   $0.0092$  &   LSQ,CRTS      &  (26)       &   P,L   \\
LSQ13cwp                        &   04:03:50.65   &   -02:39:17.98   &   2MASX J04035024-0239275          &   $0.0666$   &   LSQ      &   CSP      &   C,L   \\
LSQ13dby                        &   03:26:42.84   &   -34:38:05.49   &   anonymous                        &   $0.100$\tablenotemark{e}    &   LSQ      &   PESSTO   &   C,L   \\
LSQ13dcy                        &   04:55:16.42   &   -20:00:05.40   &   LCSB S0801P                      &   $0.0801$\tablenotemark{d}    &   LSQ      &   PESSTO   &   C,L   \\
LSQ13dhj                        &   02:12:34.48   &   -37:20:22.81   &   GALEXMSC J021234.60-372019.1     &   $0.0935$\tablenotemark{d}    &   LSQ      &   PESSTO   &   C,L   \\
LSQ13dkp                        &   03:10:09.97   &   -36:37:44.91   &   2MASX J03101094-3638017          &   $0.0690$                    &   LSQ      &   PESSTO   &   C,L   \\
LSQ13dpm                        &   10:29:08.32   &   -17:06:50.19   &   GALEXASC J102908.61-170654.2     &   $0.0509$\tablenotemark{d}   &   LSQ      &   PESSTO   &   C,L   \\
LSQ13dqh                        &   04:22:05.90   &   -02:53:24.24   &   anonymous                        &   $0.1038$\tablenotemark{d}   &   LSQ      &   PESSTO   &   C,L   \\
LSQ13dsm                        &   03:33:12.83   &   -26:12:24.02   &   APMUKS(BJ) B033105.19-262232.9   &   $0.0424$\tablenotemark{d}   &   LSQ      &   CSP      &   C,L   \\
LSQ14q                          &   08:53:57.58   &   +17:19:38.19   &   SDSS J085357.19+171942.6         &   $0.0667$                    &   LSQ,PS1      &   PESSTO   &   C,L   \\
LSQ14ba                         &   11:01:23.01   &   -15:37:10.29   &   GALEXASC J110123.17-153706.2     &   $0.0776$\tablenotemark{d}   &   LSQ      &   PESSTO   &   C,L   \\
LSQ14ie                         &   12:55:33.46   &   -32:56:40.37   &   anonymous                        &   $0.0896$\tablenotemark{d}   &   LSQ      &   PESSTO   &   C,L   \\
LSQ14ip                         &   09:44:20.22   &   +04:35:52.56   &   2MASX J09442084+0435319          &   $0.0613$                    &   LSQ      &   PESSTO   &   C,L   \\
LSQ14jp                         &   12:57:21.51   &   -15:47:33.42   &   2MASX J12572166-1547411          &   $0.0454$                    &   LSQ      &   CSP      &   C,L   \\
LSQ14mc                         &   09:02:13.45   &   +17:03:37.01   &   SDSS J090213.35+170335.4         &   $0.0567$                    &   LSQ      &   PESSTO   &   C,L   \\
LSQ14wp                         &   10:14:05.72   &   +06:40:30.54   &   SDSS J101405.83+064032.5         &   $0.0695$\tablenotemark{d}   &   LSQ,CRTS      &   PESSTO   &   C,L   \\
LSQ14xi                         &   12:30:41.17   &   -13:46:21.91   &   2MASX J12304088-1346236          &   $0.0508$                    &   LSQ      &   PESSTO   &   C,L   \\
LSQ14act                        &   15:59:44.65   &   -10:26:40.80   &   2MASX J15594429-1026396          &   $0.0591$                    &   LSQ      &   PESSTO   &   C,L   \\
LSQ14age                        &   13:24:08.62   &   -13:26:26.00   &   GALEXASC J132408.58-132629.0     &   $0.0806$\tablenotemark{d}   &   LSQ      &   PESSTO   &   C,L   \\
LSQ14ahc                        &   13:43:48.25   &   -32:54:35.12   &   2MASX J13434760-3254381          &   $0.0509$\tablenotemark{d}   &   LSQ      &   PESSTO   &   C,L   \\
LSQ14ahm                        &   11:41:22.44   &   -12:23:57.69   &   GALEXASC J114122.65-122354.9     &   $0.0498$\tablenotemark{d}   &   LSQ,CRTS      &   PESSTO   &   C,L   \\
LSQ14ajn (SN2014ah)             &   11:55:30.88   &   +11:55:25.88   &   CGCG 068-091                     &   $0.0210$                    &   LSQ,PS1      &   Asiago   &   P,L   \\
LSQ14asu                        &   11:11:35.95   &   -21:27:59.63   &   2MASX J11113635-2127597          &   $0.0684$                    &   LSQ      &   CSP      &   C,L   \\
LSQ14auy                        &   14:28:11.30   &   -04:03:17.50   &   2MASX J14281171-0403150          &   $0.0825$                    &   LSQ      &   CSP      &   C,L   \\
LSQ14azy                        &   11:12:34.73   &   +12:04:24.83   &   2MASX J11123493+1204206          &   $0.0458$                    &   LSQ      &  (27)       &   L     \\
LSQ14bbv                        &   19:59:33.12   &   -56:59:27.84   &   2MASS J19593264-5659334          &   $0.0588$\tablenotemark{d}   &   LSQ      &   CSP      &   L     \\
LSQ14bjj                        &   14:20:49.08   &   -05:15:02.25   &   APMUKS(BJ) B141811.91-050120.8   &   $0.0812$\tablenotemark{d}   &   LSQ      &   PESSTO   &   C,L   \\
LSQ14fms                        &   00:14:59.82   &   -51:12:39.54   &   2MASX J00145929-5112380          &   $0.078$\tablenotemark{d}    &   LSQ      &   CSP      &   C,L   \\
LSQ14foj                        &   00:26:34.67   &   -32:48:33.09   &   GALEXASC J002634.59-324825.5     &   $0.0461$                    &   LSQ      &   PESSTO   &   C,L   \\
LSQ14fom                        &   21:59:49.73   &   -30:16:15.56   &   2MASX J21594968-3016187          &   $0.0563$                    &   LSQ      &   PESSTO   &   C,L   \\
LSQ14gfb                        &   05:10:05.76   &   -36:18:43.57   &   2MASX J05100559-3618388          &   $0.0527$                    &   LSQ      &   PESSTO   &   C,L   \\
LSQ14gfn                        &   03:28:32.16   &   -04:12:14.22   &   2MASX J03283205-0412113          &   $0.1217$\tablenotemark{d}   &   LSQ      &   CSP      &   C,L   \\
LSQ14ghv                        &   03:23:44.15   &   -31:35:03.17   &   2MASX J03234449-3135101          &   $0.0667$                    &   LSQ      &   LCOGT    &   C,L   \\
LSQ14gov                        &   04:06:01.33   &   -16:01:41.49   &   GALEXMSC J040601.67-160139.7     &   $0.0896$\tablenotemark{d}   &   LSQ      &   PESSTO   &   C,L   \\
LSQ15bv                         &   10:59:47.29   &   -16:49:10.63   &   2MASX J10594717-1649070          &   $0.0689$                    &   LSQ      &  (28)       &   C,L   \\
LSQ15aae                        &   16:30:15.70   &   +05:55:58.73   &   2MASX J16301506+0555514          &   $0.0516$\tablenotemark{d}   &   LSQ      &   CSP      &   C,L   \\
LSQ15agh                        &   10:52:54.78   &   +23:27:41.65   &   2MASX J10525434+2335518          &   $0.0603$                    &   LSQ      &   CSP      &   C,L   \\
LSQ15aja                        &   17:03:08.92   &   +12:27:41.65   &   SDSS J170308.90+122741.5     &   $0.0700$\tablenotemark{d}   &   LSQ      &   PESSTO   &   C,L   \\
LSQ15alq                        &   13:09:18.56   &   -25:52:20.24   &   ESO 508- G 016                   &   $0.0471$                    &   LSQ,PS1      &   CSP      &   C,L   \\
\hline
MASTER OT J030559.89+043238.2   &   03:05:59.89   &   +04:32:38.20   &   SDSS J030559.63+043246.0         &   $0.0282$\tablenotemark{d}   &   MASTER   &   CfA      &   P     \\
\hline
OGLE-2012-SN-040                &   06:07:01.59   &   -69:21:17.10   &   2MASX J06070178-6921180          &   $0.0147$                    &   OGLE     &   PESSTO   &        \\
OGLE-2013-SN-015                &   02:02:21.56   &   -65:44:08.66   &   2MASX J02022241-6544090          &   $0.0890$\tablenotemark{d}   &   OGLE     &   CSP      &   C     \\
OGLE-2013-SN-109                &   01:46:09.34   &   -67:28:00.10   &   2MASX J01460987-6727579          &   $0.0868$\tablenotemark{d}    &   OGLE     &   PESSTO   &   C     \\
OGLE-2013-SN-118                &   05:14:47.50   &   -66:50:29.10   &   2MASX J05144615-6650292          &   $0.0750$\tablenotemark{d}   &   OGLE     &   PESSTO   &   C     \\
OGLE-2013-SN-123                &   05:58:30.38   &   -63:33:38.30   &   2MASX J05583036-6333386          &   $0.0614$\tablenotemark{d}   &   OGLE     &   PESSTO   &   C     \\
OGLE-2013-SN-126                &   04:19:47.21   &   -63:43:22.04   &   anonymous                        &   $0.0597$\tablenotemark{d}   &   OGLE     &   PESSTO   &   C     \\
OGLE-2013-SN-148                &   06:38:06.97   &   -75:43:37.21   &   2MASX J06380745-7543288          &   $0.0404$\tablenotemark{d}   &   OGLE     &   PESSTO   &   C     \\
OGLE-2014-SN-019                &   06:13:48.04   &   -67:55:15.00   &   2MASX J06134795-6755146          &   $0.0359$                    &   OGLE     &   PESSTO   &   C     \\
OGLE-2014-SN-021                &   05:48:23.49   &   -66:47:29.70   &   anonymous                        &   $0.0422$\tablenotemark{d}   &   OGLE     &   SMT      &   C     \\
OGLE-2014-SN-107                &   00:42:28.75   &   -64:45:51.00   &   APMUKS(BJ) B004021.02-650219.5   &   $0.0664$\tablenotemark{d}   &   OGLE     &   PESSTO   &   C     \\
OGLE-2014-SN-141                &   05:37:18.64   &   -75:43:17.00   &   2MASX J05371898-7543157          &   $0.0625$\tablenotemark{d}   &   OGLE     &   PESSTO   &   C     \\
SN2015F   &   07:36:15.76   &   -69:30:23.00   &   NGC 2442                         &   $0.0049$                    &   (2),OGLE     &   PESSTO   &   P     \\
\hline
PS1-13eao                       &   03:29:56.35   &   -28:46:17.70   &   ESO 418- G 007                   &   $0.0378$                    &   PS1      &   PESSTO   &   C     \\
PS1-14ra                        &   14:41:28.44   &   +09:25:58.70   &   IC 1044                          &   $0.0281$                    &   PS1      &   PESSTO   &   C,P   \\
PS1-14rx                        &   12:46:53.35   &   +14:47:50.10   &   SDSS J124653.32+144748.4         &   $0.0666$                    &   PS1      &   PESSTO   &   C     \\
PS1-14xw                        &   16:52:57.93   &   +02:23:36.50   &   NGC 6240                         &   $0.0245$                    &   PS1      &   PESSTO   &        \\
PS15sv                          &   16:13:11.74   &   +01:35:31.10   &   GALEXASC J161311.68+013532.2     &   $0.0333$\tablenotemark{d}   &   PS1      &   PESSTO   &   C,P   \\
SN2013ct     &   01:12:54.92   &   +00:58:45.70   &   NGC 428                          &   $0.0038$                    &   BOSS,PS1     &   CSP      &   P     \\
SN2013gy    &   03:42:16.88   &   -04:43:18.48   &   NGC 1418                         &   $0.0140$                    &   LOSS,PS1,CRTS   &   Asiago   &   P     \\
\hline
PTF11pbp (SN2011hb)             &   23:27:55.52   &   +08:46:45.00   &   NGC 7674                         &   $0.0289$                    &   PTF,CRTS      &   PTF      &   C,P   \\
PTF11ppn                        &   21:35:21.93   &   +26:56:04.70   &   2MASX J21352164+2656051          &   $0.0673$\tablenotemark{d}   &   PTF      &   PTF      &   C     \\
PTF11pra (SN2011hk)              &   02:18:45.81   &   -06:38:31.00   &   NGC 881                          &   $0.0176$                    &   PTF      &   PTF      &        \\
PTF11qnr (SN2011im)     &   22:44:25.45   &   -00:10:02.00   &   NGC 7364                         &   $0.0162$                    &   PTF      &   PTF      &   P     \\
iPTF13ez                        &   12:09:51.30   &   +19:47:15.70   &   SDSS J120951.25+194716.5         &   $0.0436$                    &   iPTF     &   iPTF     &   C   \\
iPTF13anh                       &   13:06:50.45   &   +15:34:32.36   &   SDSS J130650.44+153432.7         &   $0.0614$\tablenotemark{d}   &   iPTF,CRTS     &   iPTF     &   C     \\
iPTF13duj                       &   21:13:44.78   &   +13:34:33.10   &   NGC 7042                         &   $0.0170$                    &   iPTF     &   LCOGT    &   P     \\
iPTF13dwl                       &   21:16:49.05   &   +12:00:50.80   &   GALEXASC J211648.97+120052.8     &   $0.0875$\tablenotemark{d}   &   iPTF     &   iPTF     &   C     \\
iPTF13dym                       &   23:24:30.19   &   +14:39:04.00   &   SDSS J232430.20+143903.6         &   $0.0422$                    &   iPTF     &   iPTF     &   C     \\
iPTF13ebh                       &   02:21:59.98   &   +33:16:13.70   &   NGC 0890                         &   $0.0133$                    &   iPTF     &   CSP      &   P     \\
iPTF13efe                       &   08:43:39.30   &   +16:10:37.30   &   SDSS J084339.26+161037.5         &   $0.0751$\tablenotemark{d}   &   iPTF,CRTS     &   iPTF     &   C     \\
iPTF14fpg (SN2014dk)            &   00:28:12.00   &   +07:09:43.50   &   SDSS J002812.09+070940.0       &   $0.034$\tablenotemark{e}   &   iPTF,PS1,CRTS     &   Asiago   &   C,P   \\
iPTF14yw (SN2014aa)             &   11:45:03.58   &   +19:58:25.40   &   NGC 3861                         &   $0.0170$                    &   iPTF     &   Asiago   &   P     \\
iPTF14w                         &   12:03:31.29   &   +02:02:34.00   &   UGC 07034                        &   $0.0189$                    &   iPTF     &   Asiago   &   P     \\
iPTF14uo                        &   13:18:57.69   &   +09:42:40.30   &   GALEXASC J131857.37+094244.0     &   $0.0913$\tablenotemark{d}   &   iPTF     &   iPTF     &   C     \\
iPTF14yy                        &   12:26:09.17   &   +09:58:44.20   &   SDSS J122608.78+095847.1         &   $0.0431$                    &   iPTF     &   iPTF     &   C   \\
iPTF14aje                       &   15:25:12.07   &   -01:48:51.50   &   SDSS J152512.43-014840.1         &   $0.0276$                    &   iPTF     &   iPTF     &   C   \\
iPTF14gnl                       &   00:23:48.33   &   -03:51:27.90   &   LCSB S0066P                      &   $0.0537$                    &   iPTF     &   iPTF     &   C     \\
\hline
ROTSE3 J123935.1+163512 (SN2012G)  &   12:39:35.10   &   +16:35:11.90   &   IC 0803 NED01                    &   $0.0258$\tablenotemark{d}   &   ROTSE-III,MASTER     &  (29)       &   P     \\
\hline
SMT J03253351-5344190            &   03:25:33.51   &   -53:44:19.00   &   anonymous                        &   $0.0592$\tablenotemark{d}   &   SMT      &   SMT      &   C     \\
\hline
SN2011iv                        &   03:38:51.34   &   -35:35:32.00   &   NGC 1404                         &   $0.0065$                    &   BOSS     &  (30)       &   P     \\
SN2011iy                        &   13:08:58.38   &   -15:31:04.00   &   NGC 4984                         &   $0.0043$                    &  (3)       &  (31)      &   P     \\
SN2011jh                        &   12:47:14.42   &   -10:03:47.30   &   NGC 4682                         &   $0.0078$                    &  (4)       &   Asiago   &   P     \\
SN2012E                         &   02:33:22.79   &   +09:35:05.60   &   NGC 975                          &   $0.0203$                    &  (5)        &   Asiago   &   P     \\
SN2012U                         &   02:06:04.33   &   -55:11:37.50   &   ESO 153- G 020                   &   $0.0197$                    &   BOSS     &  (32)       &   P     \\
SN2012ah                        &   23:25:59.63   &   -81:54:33.30   &   NGC 7637                         &   $0.0124$                    &   BOSS    &  (33)      &   P     \\
SN2012bl                        &   20:23:55.28   &   -48:21:17.30   &   ESO 234-019                      &   $0.0187$                    &   CHASE    &  (34)       &   P     \\
SN2012bo                        &   12:50:45.23   &   -14:16:08.50   &   NGC 4726                         &   $0.0254$                    &  (6)        &   Asiago   &   P     \\
SN2012fr                        &   03:33:36.10   &   -36:07:34.00   &   NGC 1365                         &   $0.0055$                    &   TAROT    &   SMT      &   P     \\
SN2012gm                        &   23:17:37.03   &   +14:00:08.90   &   NGC 7580                         &   $0.0148$                    &  (7)        &   Asiago   &   P     \\
SN2012hd                        &   01:14:07.46   &   -32:39:07.70   &   IC 1657                          &   $0.0120$                    &   BOSS     &   PESSTO   &   P     \\
SN2012hr                        &   06:21:38.46   &   -59:42:50.60   &   ESO 121- G 026                   &   $0.0076$                    &   BOSS     &   CSP      &   P     \\
SN2012ht                        &   10:53:22.75   &   +16:46:34.90   &   NGC 3447                         &   $0.0036$                    &  (8)        &  (8)        &   P     \\
SN2012id                        &   04:42:41.14   &   +18:34:59.70   &   2MASX J04424248+1835003          &   $0.0157$                    &  (9)        &   Asiago   &   P     \\
SN2012ij                        &   11:40:15.84   &   +17:27:22.20   &   CGCG 097-050                     &   $0.0110$                    &   TNTS     &  (35)       &   P     \\
SN2013E                         &   10:00:05.52   &   -34:14:01.30   &   IC 2532                          &   $0.0094$                    &   BOSS     &   CSP      &   P     \\
SN2013H                         &   09:06:30.70   &   -75:49:01.50   &   ESO 036- G 019                   &   $0.0155$                    &   BOSS     &   CSP      &   P     \\
SN2013M                         &   13:59:56.68   &   -37:51:49.40   &   ESO 325- G 043                   &   $0.0350$                    &   BOSS     &  (36)       &   C,P   \\
SN2013U                         &   10:01:12.00   &   +00:19:42.30   &   CGCG 008-023                     &   $0.0345$                    &  (10)        &   Asiago   &   C,P   \\
SN2013aa                        &   14:32:33.88   &   -44:13:27.80   &   NGC 5643                         &   $0.0040$                    &   BOSS     &  (37)       &   P     \\
SN2013aj                        &   13:54:00.68   &   -07:55:43.80   &   NGC 5339                         &   $0.0091$                    &  (11)        &  (11)        &   P     \\
SN2013ay                        &   18:42:37.86   &   -64:56:13.50   &   IC 4745                          &   $0.0157$                    &   CHASE    &   CSP      &   P     \\
SN2013cg                        &   09:26:56.77   &   -24:46:59.60   &   NGC 2891                         &   $0.0080$                    &   CHASE    &  (38)       &   P     \\
SN2013fy                        &   21:37:27.12   &   -47:01:54.80   &   ESO 287- G 040                   &   $0.0309$                    &   BOSS     &   PESSTO   &   C,P   \\
SN2013fz                        &   04:23:46.44   &   -51:35:46.30   &   NGC 1578                         &   $0.0206$                    &   BOSS     &   PESSTO   &   P     \\
SN2013gv                        &   03:09:57.31   &   +19:12:49.20   &   IC 1890                          &   $0.0341$                    &  (12)      &   Asiago   &   C,P   \\
SN2013hh                        &   11:29:04.37   &   +17:14:09.50   &   UGC 06483                        &   $0.0130$                    &   TAROT    &   SMT      &   P     \\
SN2013hn                        &   13:48:59.17   &   -30:17:26.50   &   IC 4329                          &   $0.0151$                    &  (13)       &   LCOGT    &   P     \\
SN2014Z                         &   01:44:07.99   &   -61:07:07.40   &   ESO 114- G 004                   &   $0.0213$                    &   BOSS     &   CSP      &   P     \\
SN2014at                        &   21:46:14.82   &   -46:31:21.10   &   NGC7119                          &   $0.0322$                    &  BOSS    &   PESSTO   &   C,P   \\
SN2014ba                        &   22:55:01.97   &   -39:39:34.50   &   NGC 7410                         &   $0.0058$                    &   BOSS     &  (39)       &   P     \\
SN2014bz                        &   13:56:04.19   &   -43:35:09.90   &   2MASX J13560316-4334319          &   $0.0225$                    &   TAROT    &   SMT      &        \\
SN2014dn                        &   04:17:54.27   &   -56:36:45.20   &   IC 2060                          &   $0.0222$                    &   BOSS     &   CSP      &        \\
\enddata
\tablenotetext{a}{Heliocentric redshift are
from the NASA/IPAC Extragalactic Database (NED) unless otherwise indicated.}
\tablenotetext{b}{ASASSN \citep{shappee14,holoien17}; Asiago \citep{tomasella14}; BOSS (\url{http://bosssupernova.com/}); CHASE \citep{pignata09}; 
CfA (\url{https://www.cfa.harvard.edu/supernova//RecentSN.html}); CRTS \citep{djorgovski11}; CSP (this paper); iPTF \citep{kulkarni13}; 
ISSP (\url{http://italiansupernovae.org/en/project/description.html}); Gaia \citep{altavilla12}; KISS \citep{morokuma14}; LCOGT \citep{howell14}; LOSS \citep{filippenko01}; 
LSQ \citep{baltay13}; MASTER \citep{gorbovskoy13}; OGLE \citep{wyrzykowski14}; PESSTO \citep{smartt15}; PS1 \citep{kaiser10,scolnic18}; 
PTF \citep{law09}; ROTSE-III \citep{akerlof03}; SMT \citep{scalzo17}; SNF \citep{wood-vasey04}; TAROT \citep{klotz08}; TNTS \citep{yao15}}
\tablenotetext{c}{C = ``Cosmology''; P = ``Physics''; L = ``LSQ''}
\tablenotetext{d}{Redshift measured by CSP-II}
\tablenotetext{e}{Approximate redshift derived from SN spectrum}
\tablenotetext{f}{Peter Nugent (private communication)}
\tablenotetext{g}{Redshift of Coma Cluster}
\tablerefs{
(1) \citet{kangas14}; 
(2) \citet{CBET4081}; 
(3) \citet{CBET2943}; 
(4) \citet{nakano11}; 
(5) \citet{cox12}; 
(6) \citet{itagaki12}; 
(7) \citet{rich12}; 
(8) \citet{nishiyama12}; 
(9) \citet{yusa12}; 
(10) \citet{gagliano13}; 
(11) \citet{cortini13}; 
(12) \citet{kiyota13}; 
(13) \citet{kot13}; 
(14) \citet{ATel5823}  
(15) \citet{ATel6814};  
(16) \citet{ATel7086};   
(17) \citet{ATel7333};  
(18) \citet{ATel7420};  
(19) \citet{ATel4597};  
(20) \citet{CBET3335};  
(21) \citet{CBET3507}; 
(22) \citet{ATel7251}; 
(23) \citet{CBET2963}; 
(24) \citet{CBET3438}; 
(25) \citet{ATel3826}; 
(26) \citet{ATel5067}; 
(27) \citet{ATel6080}; 
(28) \citet{ATel6952}; 
(29) \citet{CBET2983}; 
(30) \citet{CBET2940,CBET2940b}; 
(31) \citet{CBET2943b,CBET2943c}; 
(32) \citet{CBET3007}; 
(33) \citet{CBET3028}; 
(34) \citet{CBET3076}; 
(35) \citet{CBET3370}; 
(36) \citet{CBET3393}; 
(37) \citet{CBET3416}; 
(38) \citet{CBET3517}; 
(39) \citet{CBET3873}
}
\end{deluxetable*}
\end{longrotatetable}

\subsection{Cosmology Subsample}
\label{sec:cosmology}

The primary goal of the CSP-II was to obtain optical and NIR light curves of a sample of at least 
100 SNe~Ia located in the smooth Hubble flow out to a redshift of $z \sim 0.1$.  These SNe, which 
comprise the CSP-II ``Cosmology'' subsample, were selected for photometric follow-up via the following 
criteria:

\begin{itemize}  

\item {\it Discovered before maximum light at optical wavelengths.}  One of the few 
disadvantages of observing SNe Ia in the NIR is that maximum brightness occurs 3--5 days before $B$ 
maximum. Hence, to ensure that NIR photometry was obtained within a few days of NIR maximum,
SNe discovered before optical maximum light were given highest priority for 
follow-up observations.

\item {\it Spectroscopically confirmed to be a normal SNe~Ia.\footnote[27]{In this paper, 
``normal'' SNe~Ia are defined to
include ``Branch-normal'' events \citep{branch93} as well as the 1991T-like \citep{filippenko92b,
phillips92} and ``transitional'' \citep{hsiao15} events which, while considered by some as extreme, 
nevertheless fall on the luminosity versus decline-rate relation for SNe~Ia \citep{burns18}.}} Surveys 
such as the Sloan Digital Sky Survey-II Supernova Survey, the Supernova Legacy Survey (SNLS), and the
Supernova Cosmology Project (SCP)
have shown that combining magnitude and color information allows an ``educated'' guess to be 
made as to the SN type
\citep{sako11,bazin11,suzuki12}.  Nevertheless, spectroscopy 
is vital for confirmation purposes and determining the approximate light-curve phase, as well 
as for eventual sub-typing.

\item{\it Discovered preferably in an untargeted search.}  A weakness of the CSP-I SNe~Ia sample is that
nearly 90\% of the events were found in targeted searches that are strongly biased toward 
luminous host galaxies \citep[e.g.,][]{kelly10}.  Hence, for the CSP-II,  preference was given 
to SNe~Ia discovered in untargeted searches to ensure that the sample was as complete and 
unbiased as possible concerning host-galaxy type, luminosity, and metallicity. 

\item{\it Host-galaxy redshift in the range $0.03 \la z \la 0.10$.}  Many of the SNe~Ia in the CSP-II
Cosmology subsample appeared in host galaxies with cataloged redshifts.  However, approximately
one third of the events were discovered in distant or low-luminosity hosts whose redshifts were unknown.
In these cases, we relied on the redshift estimated from the classification spectrum with software tools 
such as SNID \citep{blondin07}, SUPERFIT \citep{howell05}, and GELATO \citep{harutyunyan} in 
deciding whether or not to obtain follow-up photometry.  We have since obtained redshifts for nearly all
of these host galaxies (see \S\ref{sec:host_redshifts}). 

\end{itemize}

The 125 SNe~Ia comprising the Cosmology subsample cover a redshift range of $0.027 < z < 0.137$,
with a median redshift of $z = 0.056$.  Table~\ref{tab:sources} summarizes the
various supernova surveys from which these SNe were drawn.  Fully 96\% came from untargeted 
searches, with nearly half (48\%) of the subsample having been discovered by the La Silla-QUEST Low Redshift 
Supernova Survey \citep{baltay13}.  The individual SNe belonging to the Cosmology subsample are identified 
in the final column of Table~\ref{tab:allsne} by the letter ``C''.   

\begin{deluxetable}{lccc}
\tabletypesize{\scriptsize}
\tablecolumns{4}
\tablewidth{0pt}
\tablecaption{CSP-II Cosmology Subsample Sources\label{tab:sources}}
\tablehead{
\colhead{Source} &
\colhead{Untargeted?} &
\colhead{\# of SNe} &
\colhead{Percentage}
}
\startdata
LSQ\tablenotemark{a}          & Yes & 60 & 48.0\% \\
CRTS\tablenotemark{b}       & Yes & 15 & 12.0\% \\
ASASSN\tablenotemark{c}      & Yes & 14 & 11.2\% \\
PTF/iPTF\tablenotemark{d}  & Yes & 12 & 9.6\% \\
OGLE\tablenotemark{e}       & Yes & 10 & 8.0\% \\
PS1\tablenotemark{f}           & Yes & 4 & 3.2\% \\
KISS\tablenotemark{g}         & Yes & 3 & 2.4\% \\
SMT\tablenotemark{h}          & Yes & 1 & 0.8\% \\
MASTER\tablenotemark{i}   & Yes & 1 & 0.8\% \\
 & & {---}{---} & {---}{---} \\
 & & 120 & 96.0\% \\
 & & & \\
 BOSS\tablenotemark{j}       & No & 3 & 2.4\% \\
Other\tablenotemark{l}         & No & 2 & 1.6\% \\
 & & {---}{---} & {---}{---} \\
 & & 5 & 4.0\% \\
\enddata
\tablenotetext{a}{La Silla-Quest Low Redshift Survey \citep{baltay13}}
\tablenotetext{b}{Catalina Real Time Transit Survey \citep{djorgovski11}}
\tablenotetext{c}{All-Sky Automated Survey for SuperNovae \citep{shappee14,holoien17}}
\tablenotetext{d}{Palomar Transient Factory \citep{law09} /Intermediate Palomar Transient Factory \citep{kulkarni13}}
\tablenotetext{e}{OGLE-IV Real-Time Transient Search \citep{wyrzykowski14}}
\tablenotetext{f}{Pan-STARRS1 Medium Deep Survey \citep{kaiser10,scolnic18}}
\tablenotetext{g}{Kiso Supernova Survey \citep{morokuma14}}
\tablenotetext{h}{SkyMapper Transient Survey \citep{scalzo17}}
\tablenotetext{i}{Mobile Astronomy System of TElescope Robots \citep{gorbovskoy13}}
\tablenotetext{j}{Backyard Observatory Supernova Search \url{http://bosssupernova.com/}}
\tablenotetext{k}{Italian Supernovae Search Project \url{http://italiansupernovae.org/en/project/description.html}}
\tablenotetext{l}{\citet{gagliano13,kiyota13}}
\end{deluxetable}

\begin{figure}[h]
\epsscale{0.8}
\plotone{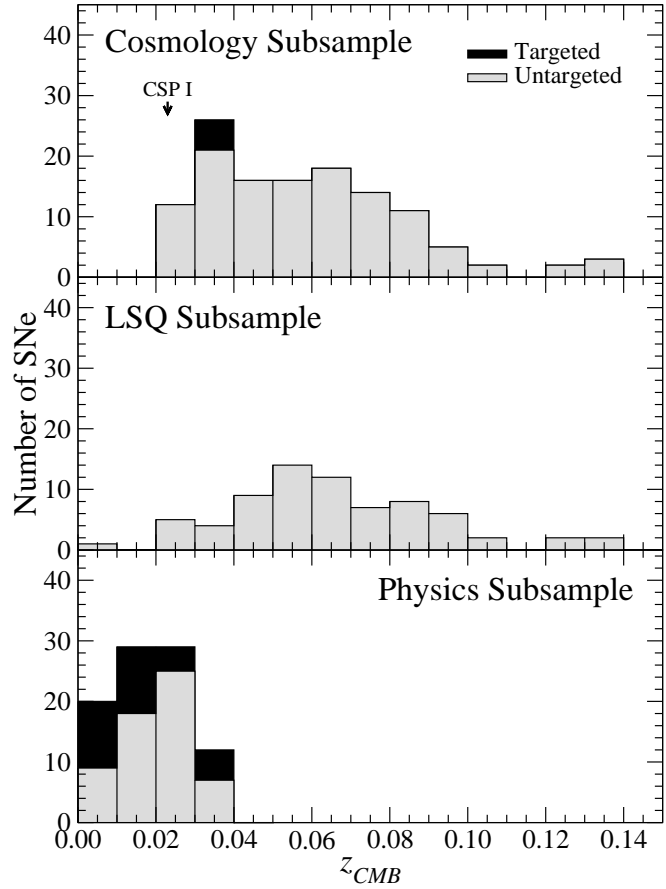} 
\caption{(Top) Histogram of heliocentric redshifts of the 125 SNe~Ia comprising the Cosmology
subsample, and (bottom) the 90 SNe~Ia in the Physics subsample.  In the top panel, the median redshift
of the CSP-I  sample is indicated by an arrow.  The middle panel displays a histogram of the redshifts of
the LSQ subsample.}
\label{fig:redshifts}
\end{figure}

\begin{figure}[h]
\epsscale{0.8}
\plotone{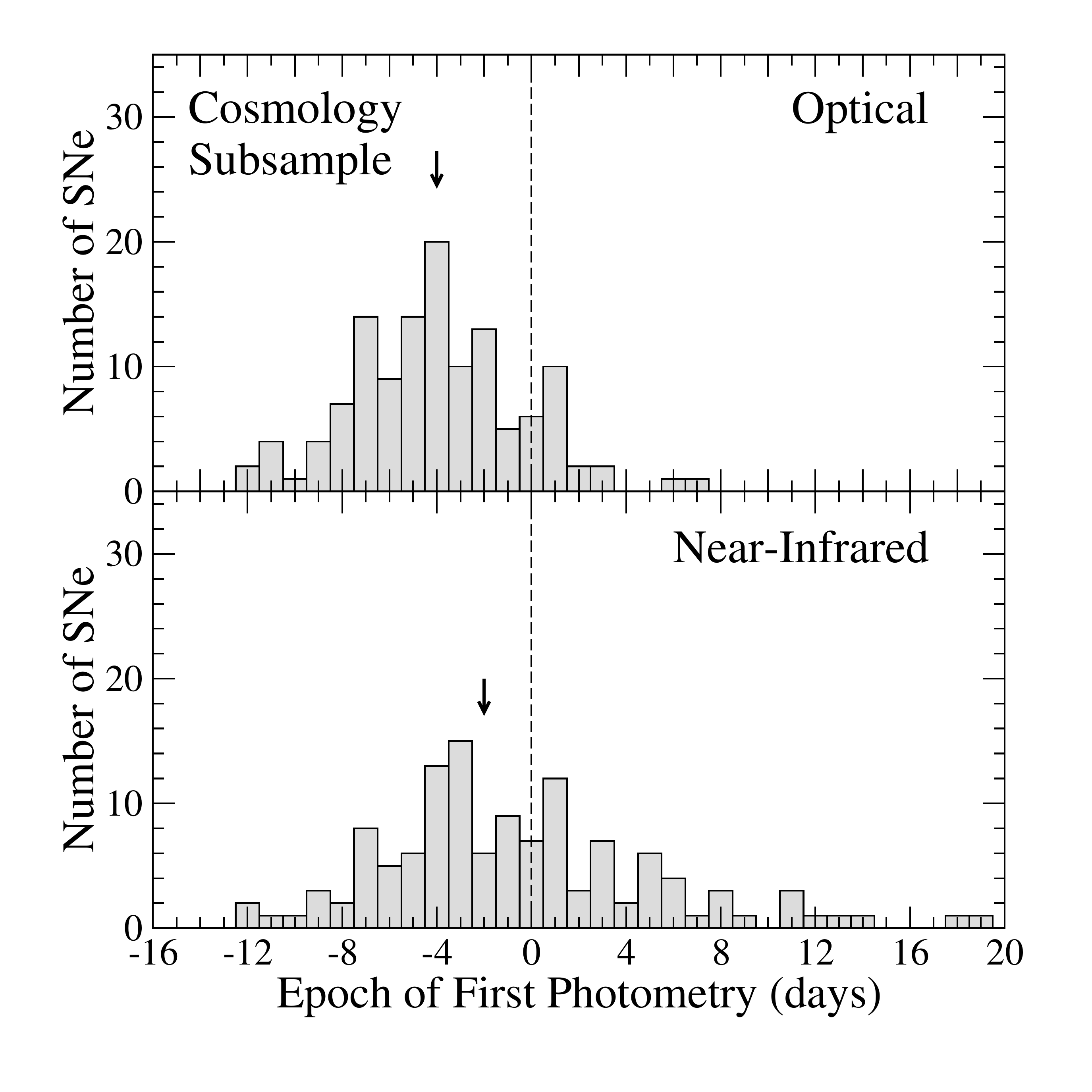} 
\caption{(Top) Histogram of the epoch with respect to the date of $B$ maximum of the first optical imaging
observations of the 125 SNe~Ia comprising the Cosmology subsample, and (bottom) similar histogram
for the first NIR imaging observations.  The arrows in each panel indicate the medians of the histograms
corresponding to $-4$ days for the first optical photometry and $-2$ days for the first NIR observation.}
\label{fig:opt_ir_start}
\end{figure}

The top panel of Figure~\ref{fig:redshifts} shows a histogram of the heliocentric redshifts of the
Cosmology subsample.  The arrow indicates the median redshift of the CSP-I sample. 
In Figure~\ref{fig:opt_ir_start}, histograms of the epoch
with respect to the date of $B$ maximum of the first imaging observations in the optical and the NIR
for the Cosmology subsample are displayed.  As may be seen, optical imaging was obtained
for half of the subsample at $-4$~days or earlier, and $-2$~days or earlier in the NIR.

\subsection{Physics Subsample}
\label{sec:physics}

To realize the full potential of SNe~Ia as distance indicators at NIR wavelengths, we must 
determine accurate K-corrections, which account for the effect of cosmological expansion upon the measured 
magnitudes \citep{oke68}. Poorly understood K-corrections directly impact the peak magnitudes of the SNe 
and inflate both statistical and systematic errors.  Prior to the CSP-II, NIR spectra had been published for
only 33 SNe~Ia, with the total number of useful spectra amounting to 75.  \citet{boldt14} used this 
sample to study the errors inherent in NIR K-corrections.  Their main finding was that uncertainties 
due to the diversity of spectral features from object to object are the dominant source of error at 
maximum light.  \citeauthor{boldt14} demonstrated that, with the small number of spectra in hand,
K-correction uncertainties in the $Y$, $J$, and $H$ bands amounted to 0.04, 0.06, and 0.10~mag 
for a SN~Ia at $z = 0.08$.  For a sample of 59 spectra of 10 SNe~Ia --- 35 spectra of which 
overlapped with the \citeauthor{boldt14} sample --- \citet{stanishev18} recently found somewhat
smaller dispersions of 0.03 and 0.04 mag in the $J$ and $H$ K-corrections at $z = 0.08$.
Much of the diversity in the spectral features is known to correlate with the 
light-curve decline rate \citep{hsiao00}, but to quantify this requires a much larger NIR spectral sample.

To attack the NIR K-correction problem, the CSP-II, in conjunction with the Harvard-Center for Astrophysics 
Supernova Group, initiated in 2012 a program of NIR spectroscopic observations of a large ``Physics'' 
subsample\footnote[28]{We choose to refer to these SNe as the ``Physics'' subsample since, as described
in detail in \citet{hsiao18}, the NIR spectral data  are also an invaluable resource for studying the 
physics and the progenitors of SNe Ia.} of nearby SNe~Ia at $z \leq 0.04$.  This collaboration is 
described in detail in an accompanying paper \citep{hsiao18}.  In order to quantify the light-curve 
properties of this subsample, it was necessary to obtain optical imaging with a cadence and 
signal-to-noise ratio comparable to (or better than)
that obtained for the Cosmology subsample.  Although imaging in the NIR was not a requirement for the Physics 
subsample, it was nonetheless obtained for 89\% of the SNe. The same basic criteria used for the 
Cosmology subsample were also applied in selecting the SNe to be included in the Physics subsample, 
although at these lower redshifts, a higher fraction (39\%) of the events were discovered in targeted 
searches.  The Physics subsample also includes several fast-declining, SN~1991bg-like 
\citep{filippenko92a,leibundgut93,ruiz-lapuente93} events that, because of their low luminosities, are 
not well represented in the Cosmology subsample.

\begin{deluxetable}{lccc}
\tabletypesize{\scriptsize}
\tablecolumns{4}
\tablewidth{0pt}
\tablecaption{CSP-II Physics Subsample Sources\label{tab:sources_physics}}
\tablehead{
\colhead{Source} &
\colhead{Untargeted?} &
\colhead{\# of SNe} &
\colhead{Percentage}
}
\startdata
ASASSN\tablenotemark{a}      & Yes & 27 & 30.0\% \\
LSQ\tablenotemark{b}          & Yes & 9 & 10.0\% \\
CRTS\tablenotemark{c}       & Yes & 8 & 8.9\% \\
PTF/iPTF\tablenotemark{d}  & Yes & 7 & 7.8\% \\
PS1\tablenotemark{e}           & Yes & 4 & 4.4\% \\
KISS\tablenotemark{f}         & Yes & 1 & 1.1\% \\
MASTER\tablenotemark{g}   & Yes & 1 & 1.1\% \\
OGLE\tablenotemark{h}   & Yes & 1 & 1.1\% \\
ROTSE-III\tablenotemark{i}  & Yes & 1 & 1.1\%\\
 & & {---}{---} & {---}{---} \\
 & & 59 & 65.6\% \\
 & & & \\
BOSS\tablenotemark{j}       & No & 14 & 15.6\% \\
CHASE\tablenotemark{k}          & No & 3 & 3.3\% \\
TAROT\tablenotemark{l}          & No & 2 & 2.2\% \\
TNTS\tablenotemark{m}          & No & 1 & 1.1\% \\
Other\tablenotemark{n}         & No & 11 & 12.2\% \\
 & & {---}{---} & {---}{---} \\
 & & 31 & 34.4\% \\
\enddata
\tablenotetext{a}{All-Sky Automated Survey for SuperNovae \citep{shappee14,holoien17}}
\tablenotetext{b}{La Silla-Quest Low Redshift Survey \citep{baltay13}}
\tablenotetext{c}{Catalina Real Time Transit Survey \citep{djorgovski11}}
\tablenotetext{d}{Palomar Transient Factory \citep{law09} /Intermediate Palomar Transient Factory \citep{kulkarni13}}
\tablenotetext{e}{Pan-STARRS1 Medium Deep Survey \citep{kaiser10,scolnic18}}
\tablenotetext{f}{Kiso Supernova Survey \citep{morokuma14}}
\tablenotetext{g}{Mobile Astronomy System of TElescope Robots \citep{gorbovskoy13}}
\tablenotetext{h}{OGLE-IV Real-Time Transient Search \citep{wyrzykowski14}}
\tablenotetext{i}{Robotic Optical Transient Search Experiment III \citep{akerlof03}}
\tablenotetext{j}{Backyard Observatory Supernova Search \url{http://bosssupernova.com/}}
\tablenotetext{k}{CHilean Automatic Supernova sEarch \citep{pignata09}}
\tablenotetext{l}{T\'elescopes \`a Action Rapide pour les Objets Transitoires \citep{klotz08}}
\tablenotetext{m}{THU-NAOC Transient Survey \citep{zhang15}}
\tablenotetext{o}{\citet{CBET2943,nakano11,cox12,itagaki12,rich12,nishiyama12,yusa12,gagliano13,cortini13,kiyota13,kot13}}
\end{deluxetable}

A total of 90 SNe~Ia with a median redshift of $z = 0.021$ comprise the Physics subsample.  
Table~\ref{tab:sources_physics} provides information on the supernova surveys from which these 
SNe were drawn, and Table~\ref{tab:allsne} lists the individual SNe (identified by the letter ``P''
in the final column).  Their distribution as a function of redshift is plotted in the bottom panel of 
Figure~\ref{fig:redshifts}.  Note that there is some overlap between the Physics and Cosmology 
subsamples, with 21 of the SNe (23\%) in the Physics subsample also forming part of the Cosmology subsample.

\subsection{La Silla-QUEST Subsample}
\label{sec:lsq}

The La Silla-QUEST supernova survey was fundamental to the success of the CSP-II,
contributing nearly half of the SNe~Ia making up the Cosmology subsample.  The LSQ subsample 
is particularly important since it was an untargeted search with homogeneous selection criteria. 
The recent dark energy analysis of the Pantheon sample of SNe~Ia by \citet{scolnic18} found that 
systematic errors are still a serious problem, particularly in modeling the low redshift sample 
because of the uncertainty of whether it is volume or magnitude limited. Calibration errors, which 
in the Pantheon sample are twice as large for the low-redshift events compared to the high-redshift 
SNe, also continue to be a significant additional source of systematic error.  The LSQ subsample 
observed by the CSP-II addresses both of these issues.

The CSP-II obtained light curves of a total of 72 SNe~Ia discovered by the LSQ survey, spanning 
a redshift range of $0.009 \leq z \leq 0.137$.  Preliminary photometry of 31 of these SNe was 
published by \citet{walker15}.  The LSQ subsample is identified in Table~\ref{tab:allsne} by the 
letter ``L'' in the last column.  Their redshift distribution is plotted in the middle panel of 
Figure~\ref{fig:redshifts}. Note that 83\% of the LSQ events are also members of the 
Cosmology subsample, and 13\% overlap with the Physics subsample.  NIR photometry
was obtained for 85\% of the LSQ subsample.

\subsection{Other SNe~Ia}
\label{sec:other_Ia}

Table~\ref{tab:allsne} also includes 11 SNe~Ia for which photometric observations were obtained,
but which do not fit into any of the three aforementioned subsamples.  These objects are 
identified by a blank entry in the ``Subsample'' column of the table.  All have redshifts too small 
($z \leq 0.026$) to be included in the Cosmology subsample, and since 
NIR spectroscopy was not obtained they do not qualify for the Physics subsample.  However, the 
light curves for these SNe are still of value, and so we include them in the full sample of CSP-II SNe Ia.

\subsection{Other Types of SNe}
\label{sec:other}

The emphasis of the CSP-II was on observing the light curves and spectra of normal SNe~Ia. Nevertheless, 
data were obtained for other types, including peculiar SNe~Ia (five SN~2002cx-like events
and four possible ``super-Chandrasekhar'' SNe), four super-luminous SNe, and several core-collapse 
and stripped core-collapse events.  The CSP-II observations of these SNe will be presented in future
papers.

\section{Observing Strategy}
\label{sec:strategy}

\subsection{Optical Imaging}
\label{sec:opt_imaging}

As was the case for the CSP-I, nearly all of the optical imaging during the CSP-II was obtained with the 
LCO Swope telescope.  Generally speaking, for redshifts less than $\la 0.04$, the same complement 
of $ugriBV$ filters used by the CSP-I was also employed.  For redshifts greater than this, a subset 
consisting of the $BVri$ filters was normally used.  Each SN was typically observed every 2--3 days from 
discovery until at least 2--3 weeks past maximum to sample the light-curve maxima and early decline 
rates as thoroughly as possible.  Optical imaging of candidate SNe was often initiated before a classification 
spectrum was obtained.  A quick reduction of the photometry was made the morning after each 
observation, allowing a nearly ``real-time'' update of the light curves on the CSP-II webpage, with the calibration
improving as data were obtained on more photometric nights.  If spectral observations subsequently 
showed that the target was not a SN~Ia, we usually discontinued the optical imaging.

As was also done for the CSP-I, the LCO 2.5~m du~Pont telescope was used to obtain most of the
host-galaxy reference images in $ugriBV$.  This telescope was also used to obtain a small amount of  
imaging of SNe being actively followed during scheduled nights.

\subsection{NIR Imaging}
\label{sec:nir_imaging}

SNe~Ia are excellent standard candles in the NIR when observed at maximum light 
\citep[e.g.,][]{krisciunas04a,kattner12}.  When combined with optical photometry, NIR 
observations at maximum also afford the most precise measurement of the host-galaxy dust reddening 
\citep{krisciunas00,mandel11,burns14,burns18}.  While the strength of the prominent NIR secondary 
maximum is a strong function of the decline rate \citep{hamuy96b}, there is significant scatter in
the correlation \citep{krisciunas01,burns14}.  
Hence, to ensure the highest precision for measuring both the host extinction and 
distance from optical and NIR observations, it is best that NIR photometry be obtained 
within $\sim$1~week of optical ($B$-band) maximum if at all possible \citep{krisciunas04b}.
As shown in Figure~\ref{fig:opt_ir_start}, only a small fraction (11/125) of the SNe in the 
Cosmology subsample do not meet this condition.

The SNe in the CSP-I sample were sufficiently nearby and bright enough that all of the optical and 
most of the NIR imaging could be obtained with the Swope telescope.  However, extending 
NIR observations to $z \sim 0.1$ requires a larger telescope, and so a more 
economical approach was necessary for the CSP-II.  We requested approximately one week of 
NIR imaging each bright run with the du~Pont 2.5~m telescope during the CSP-II observing
campaigns to cover the primary maxima of $\sim$5 SNe Ia per bright run.  For SNe 
at $z \la 0.07$, imaging was generally obtained in the $Y$, $J$, and $H$ bands, while for more 
distant targets, observations were typically restricted to the $Y$ and $J$ filters.  In a few cases 
($<$~10\%), only the $Y$ filter was used.  At least four epochs of NIR imaging were obtained
for 66\% of the SNe in the Cosmology subsample, and at least three epochs were obtained for 82\%.  
Sparse sampling of the NIR maximum of SNe~Ia has been successfully employed
in the past by \citet{krisciunas04a}, \citet{freedman09}, \citet{barone-nugent12}, and
\citet{weyant14,weyant18}.

Although most of the NIR photometry for the CSP-II was acquired with the du~Pont telescope,
we obtained a few epochs of additional imaging (1--2 epochs) with the LCO Magellan 
Baade 6.5~m telescope for nine SNe~Ia.  Many of the NIR host galaxy images were also
acquired with Magellan Baade.

\subsection{Optical Spectroscopy}
\label{sec:spectra}

Optical spectroscopy near maximum light provides an essential tool for characterizing 
the diversity of SNe~Ia which, in turn, is related to the progenitor systems and explosion 
mechanism \citep[e.g.,][]{blondin12,folatelli13}.  In total, 308 optical spectra were obtained 
of more than 100 of the SNe~Ia in the Cosmology and Physics subsamples.
These spectra were acquired with the du~Pont and
Magellan telescopes at LCO and with the 2.5~m Nordic Optical Telescope (NOT) at the Observatorio 
del Roque de los Muchachos with the principal aim of determining the type, phase, and approximate 
redshift of the SN targets.  Through these observations and the NIR spectroscopy described in
\S\ref{sec:nir_spectra} and \citet{hsiao18}, the CSP-II was able to classify $\sim$20\% of the SNe 
in both the Cosmology and Physics subsamples (listed as ``CSP'' in the Classification column of 
Table~\ref{tab:allsne}).  Many
of the remaining targets were classified by the Public ESO Spectroscopic Survey of Transient 
Objects (PESSTO) with the ESO La Silla 3.6~m NTT \citep{smartt15}. 
With the CSP-II, PESSTO, and other
publicly-available spectra, 114 of the SNe in the Cosmology and Physics subsamples were
observed within $\pm4$~days of maximum light.  
These data will be presented and analyzed in a future paper (Morrell et al., in preparation).

\subsection{Optical Integral Field Spectroscopy}
\label{sec:ifu}

The pioneering work of the Cal\'{a}n/Tololo Project showed that SNe~Ia luminosities are correlated 
with host-galaxy morphologies and colors \citep{hamuy95,hamuy96a,hamuy00}.  More recently, 
evidence has been presented that SNe~Ia Hubble diagram residuals correlate with global host-galaxy
parameters such as total mass and star formation rate \citep{kelly10,lampeitl10,sullivan10,rigault13,uddin17}.  
In order to examine these effects
for the CSP-II sample of SNe~Ia,  a program of integral field unit (IFU) host-galaxy spectroscopy is being 
carried out with both the Multi Unit Spectroscopic Explorer (MUSE) on the ESO VLT \citep{bacon10} as part of the 
All-weather MUse Supernova Integral-field Nearby Galaxies (AMUSING) survey \citep{galbany16}, and
the Potsdam Multi-Aperture Spectrophotometer (PMAS) on the 3.5m Calar Alto telescope within 
the PMAS/PPak Integral-field Supernova hosts COmpilation (PISCO) program \citep{galbany18} for those 
targets in the northern hemisphere.
In addition to the global host properties, these IFU data will provide spectral information 
on the immediate environments of the SNe --- e.g., line-of sight gas-phase and stellar metallicity, 
stellar age, and star-formation rates --- allowing a detailed
study of the correlations between the SNe~Ia Hubble diagram residuals and 
the local environmental properties.

\subsection{NIR Spectroscopy}
\label{sec:nir_spectra}

NIR spectroscopy was a new and vital component of the CSP-II.  During the four campaigns,
NIR spectra were obtained of 157 different SNe Ia.  These observations were carried out using the 
Folded-port IR Echellette (FIRE) on the LCO Magellan Baade telescope and through target of 
opportunity (ToO) time obtained principally with the Gemini North and South 8.1~m telescopes, the NASA 
Infrared Telescope Facility (IRTF), and the ESO Very Large Telescope (VLT). This set of NIR 
spectra is more than 15 times 
larger than the previous largest sample \citep{marion09}.  The FIRE spectra account for 80\% of the 
total spectra obtained, but the ToO observations were crucial for obtaining spectral coverage at the 
earliest epochs. More than 70\% of the SNe Ia observed have at least 3 epochs of NIR spectral 
observations, and more than 10 epochs were obtained for 15 SNe. Whenever possible, we also 
attempted to obtain simultaneous optical spectroscopy to match the NIR observations.
The NIR spectroscopy part of the CSP-II is presented in detail
in the accompanying paper by \citet{hsiao18}.

\subsection{Host-Galaxy Redshifts}
\label{sec:host_redshifts}

The majority of the SNe~Ia in both the Cosmology and Physics subsamples were discovered in 
untargeted searches.  Approximately one-third of the events selected for follow-up appeared in 
distant or low-luminosity hosts whose redshifts were unknown.  A program of host-galaxy 
spectroscopy was initiated to measure redshifts for these SNe. Spectra of 40 of the hosts 
were obtained with the Wide-Field CCD (WFCCD) spectrograph on the du~Pont telescope.  
For the faintest galaxies, it was necessary to use the Magellan telescopes.  Spectra of 19 hosts were 
taken with the Inamori Magellan Areal Camera and Spectrograph (IMACS) on the Baade telescope, 
and another 5 galaxies were observed with the Low-Dispersion Survey Spectrograph (LDSS3-C) 
on the Clay telescope.  Redshifts were obtained for an additional 11 host galaxies from our optical 
integral field spectroscopy program. 

The resulting redshifts are listed in Table~\ref{tab:redshifts}, which gives 
the number of absorption and emission lines used to calculate the redshifts for each host-galaxy 
spectrum, and the rms errors of the final redshift values.  These redshifts are also included in 
Table~\ref{tab:allsne}.
Note that there are eight galaxies in Table~\ref{tab:allsne} for which a redshift has not yet
been measured.

\begin{longrotatetable}
\startlongtable
\begin{deluxetable*}{llclcc}
\tabletypesize{\scriptsize}
\tablecolumns{5}
\tablewidth{0pt}
\tablecaption{CSP-II Measurements of Host Galaxy Redshifts\label{tab:redshifts}}
\tablehead{
 & & & &  \multicolumn{2}{c}{lines\tablenotemark{c}}  \\
\colhead{SN} &
\colhead{Host Galaxy} &
\colhead{Instrument(s)\tablenotemark{a}} &
\colhead{$z_{Helio}$\tablenotemark{b}} &
\colhead{a} &
\colhead{e} 
}
\startdata
ASASSN-14hu & ESO 058- G 012                   & 1 & $0.0216 \pm 0.0001$ & 1 & 5 \\ 
ASASSN-14jz & GALEXASC J184443.33-524819.2     &  1 & $0.0158 \pm 0.0001$ & \nodata & 8 \\
ASASSN-14kd & 2MASX J22532475+0447583          & 2 & $0.0242 \pm 0.0002$ & 1 & 3 \\
ASASSN-14lw & GALEXASC J010647.95-465904.1     &  2 & $0.0209 \pm 0.0001$ & \nodata & 3 \\
ASASSN-14me & ESO 113- G 047                   &   4 & $0.0178 \pm 0.0001$ & \nodata & 5  \\
ASASSN-15al & GALEXASC J045749.46-213526.3     &  1 & $0.0338 \pm 0.0001$ & 2 & 3 \\
ASASSN-15as & SDSS J093916.69+062551.1         &  1 & $0.0286 \pm 0.0001$ & \nodata & 3 \\
ASASSN-15bm & LCRS B150313.2-052600            &  1 & $0.0208 \pm 0.0003$ & 1 & 4 \\
ASASSN-15gr & ESO 366- G 015                   &  1 & $0.0243 \pm 0.0002$ & \nodata & 5 \\
ASASSN-15hx & GALEXASC J134316.80-313318.2     &  4 & $0.0083 \pm 0.0002$ & \nodata & 4 \\
CSS120325:123816-150632 & anonymous                        &  2 & $0.0972 \pm 0.0002$ & \nodata & 5 \\
CSS130215:033841+101827 (SN2013ad) & anonymous                        &  2 & $0.0363 \pm 0.0001$ & \nodata & 4 \\
CSS130303:105206-133424 & GALEXASC J105206.27-133420.2     &  1 & $0.0789 \pm 0.0004$ & \nodata & 1 \\
CSS130315:115252-185920 (SN2013as) & anonymous                        & 1 & $0.0685 \pm 0.0001$ & \nodata & 7 \\
CSS140126:120307-010132         &   SDSS J120306.76-010132.4 & 2 & $0.0772 \pm 0.0001$ & \nodata & 4 \\
CSS140218:095739+123318 & SDSS J095738.31+123308.5         & 2 & $0.0773 \pm 0.0002$ & \nodata & 1 \\
LSQ11bk  & anonymous                        &  2 & $0.0403 \pm 0.0001$ & \nodata & 5 \\
LSQ12ca  & 2MASX J05310364-1948063          &  4 & $0.0994 \pm 0.0003$ & \nodata & 3  \\
LSQ12agq & GALEXASC J101741.80-072452.2     &  3 & $0.0642 \pm 0.0001$ & \nodata & 6 \\
LSQ12aor & GALEXASC J105517.85-141757.2     &  3 & $0.0934 \pm 0.0001$ & \nodata & 8 \\
LSQ12cdl & GALEXASC J125339.85-183025.6     &  2 & $0.1081 \pm 0.0001$ & \nodata & 6 \\
LSQ12fuk & GALEXASC J045815.88-161800.7     & 1 & $0.0206 \pm 0.0002$ & \nodata & 4 \\
LSQ12gef & 2MASX J01403375+1830406          & 1 & $0.0642 \pm 0.0004$ & 6 & 3 \\
LSQ12gln & GALEXASC J052259.58-332755.3     & 3 & $0.1021 \pm 0.0001$ & \nodata & 8 \\
LSQ12gpw & 2MASX J03125885-1142402          &  1 & $0.0506 \pm 0.0002$ & 1 & 4 \\
LSQ12gzm & GALEXASC J024043.58-344425.0     &  4 & $0.1001 \pm 0.0002$ & \nodata & 5  \\
LSQ12hjm & 2MASX J03102844-1629333          & 1 & $0.0714 \pm 0.0002$ & 2 & 6 \\
LSQ12hno & GALEXASC J034243.43-024007.7     &  1 & $0.0473 \pm 0.0001$ & \nodata & 7 \\
LSQ12hvj & GALEXASC J110738.65-294235.5     & 1 & $0.0713 \pm 0.0001$ & \nodata & 4 \\
LSQ12hzj & 2MASX J09591230-0900095          &  2,4 & $0.0334 \pm 0.0002$ & 4 & \nodata \\
LSQ13lq & SDSS J134410.77+030345.3         & 1 & $0.0757 \pm 0.0002$ & \nodata & 4 \\
LSQ13pf & LCRS B134534.3-112338            & 2 & $0.0861 \pm 0.0001$ & 1 & 3 \\
LSQ13vy  & 2MASX J16065563+0300046          &  1 & $0.0418 \pm 0.0004$ & 4 & 4 \\
LSQ13dcy & LCSB S0801P                      &  1,4 & $0.0801 \pm 0.0003$ & 7 & \nodata \\
LSQ13dhj & GALEXMSC J021234.60-372019.1     &  1,4 & $0.0935 \pm 0.0001$ & \nodata & 5 \\
LSQ13dpm & GALEXASC J102908.61-170654.2     & 1 & $0.0509 \pm 0.0003$ & \nodata & 4 \\
LSQ13dqh & anonymous                        &  2 & $0.1038 \pm 0.0001$ & \nodata & 4 \\
LSQ13dsm & APMUKS(BJ) B033105.19-262232.9   &  1 & $0.0424 \pm 0.0002$ & \nodata & 5 \\
LSQ14ba & GALEXASC J110123.17-153706.2     & 1 & $0.0776 \pm 0.0001$ & \nodata & 4 \\
LSQ14ie  & anonymous                        & 1 & $0.0896 \pm 0.0001$ & \nodata & 5 \\
LSQ14wp & SDSS J101405.83+064032.5         &  2 & $0.0695 \pm 0.0001$ & \nodata & 4 \\
LSQ14age & GALEXASC J132408.58-132629.0     &  1 & $0.0806 \pm 0.0003$ & \nodata & 5 \\
LSQ14ahc & 2MASX J13434760-3254381          & 2 & $0.0509 \pm 0.0001$ & \nodata & 3 \\
LSQ14ahm & GALEXASC J114122.65-122354.9     & 1 & $0.0498 \pm 0.0001$ & \nodata & 4 \\
LSQ14bbv & 2MASS J19593264-5659334          &  1 & $0.0588 \pm 0.0006$ & 7 & \nodata \\
LSQ14bjj & APMUKS(BJ) B141811.91-050120.8   & 1 & $0.0812 \pm 0.0001$ & \nodata & 7 \\
LSQ14fms & 2MASX J00145929-5112380          & 1 & $0.078 \pm 0.001$  & 3 & 2 \\
LSQ14gfn & 2MASX J03283205-0412113          &  2 & $0.1217 \pm 0.0004$ & 6 & \nodata \\ 
LSQ14gov & GALEXMSC J040601.67-160139.7     & 1 & $0.0896 \pm 0.0002$ & \nodata & 5 \\
LSQ15aae & 2MASX J16301506+0555514          & 1 & $0.0516 \pm 0.0001$ & 1 & 2 \\
LSQ15aja & SDSS J170308.90+122741.5     &  2 & $0.0700 \pm 0.0001$ & \nodata & 4 \\
MASTER OT J030559.89+043238.2 & SDSS J030559.63+043246.0         &  1 & $0.0282 \pm 0.0009$ & \nodata & 3 \\
OGLE-2013-SN-015 & 2MASX J02022241-6544090          &  1,4 & $0.0890 \pm 0.0002$ & 4 & \nodata \\
OGLE-2013-SN-109 & 2MASX J01460987-6727579          & 1,4 & $0.0868 \pm 0.0001$ & \nodata & 5 \\
OGLE-2013-SN-118 & 2MASX J05144615-6650292          &  1 & $0.0750 \pm 0.0002$ & \nodata & 5 \\
OGLE-2013-SN-123 & 2MASX J05583036-6333386          & 1 & $0.0614 \pm 0.0002$ & 9 & \nodata \\
OGLE-2013-SN-126 & anonymous                        &  2 & $0.0597 \pm 0.0002$ & \nodata & 3 \\
OGLE-2013-SN-148 & 2MASX J06380745-7543288          & 1 & $0.0404 \pm 0.0002$ & \nodata & 6 \\
OGLE-2014-SN-021 & anonymous                        &  3 & $0.0422 \pm 0.0002$ & \nodata & 6 \\
OGLE-2014-SN-107 & APMUKS(BJ) B004021.02-650219.5   & 1 & $0.0664 \pm 0.0001$ & \nodata & 3 \\
OGLE-2014-SN-141 & 2MASX J05371898-7543157          & 1 & $0.0625 \pm 0.0002$ & 1 & 8 \\ 
PS15sv   & GALEXASC J161311.68+013532.2     &  1 & $0.0333 \pm 0.0003$ & \nodata & 5 \\ 
PTF11ppn & 2MASX J21352164+2656051          & 1 & $0.0673 \pm 0.0002$ & 8 & \nodata \\
iPTF13anh & SDSS J130650.44+153432.7         &  2 & $0.0614 \pm 0.0001$ & \nodata & 6 \\
iPTF13dwl & GALEXASC J211648.97+120052.8     & 1 & $0.0875 \pm 0.0002$ & \nodata & 6 \\
iPTF13efe & SDSS J084339.26+161037.5         & 2 & $0.0751 \pm 0.0001$ & \nodata & 4 \\
iPTF14uo & GALEXASC J131857.37+094244.0     & 2 & $0.0913 \pm 0.0001$ & \nodata & 7 \\
ROTSE3 J123935.1+163512 & IC 0803 NED01 (SN2012G)      &  4 & $0.0258 \pm 0.0002$ & \nodata & 5  \\
SMTJ03253351-5344190 & anonymous                        &  3 & $0.0592 \pm 0.0002$ & \nodata & 5 \\
SNhunt161 (SN2012hl)   & CSS J005017.69+243154.4  & 5 & $0.0332 \pm 0.0002$ & \nodata & 6  \\
\enddata
\tablenotetext{a} {1: du Pont+WFCCD, 2: Magellan Baade+IMACS, 3: Magellan Clay+LDSS3, 4: VLT+MUSE, 5: Calar Alto~3.5~m+PMAS/PPak}
\tablenotetext{b} {The heliocentric redshift measurement was derived from the equivalent-width-weighted average of 
the features listed in final column.  A minimum error of 0.0001 is adopted from instrumental resolution considerations.}
\tablenotetext{c} {Number of absorption (a) and emission (e) lines used to measure the redshift.  Wavelengths of the
features used to calculate the redshift are taken from Table~II of \citet{sandage78}.}
\end{deluxetable*}
\end{longrotatetable}

\section{Optical Photometry}
\label{sec:opt_phot}

\subsection{Swope 1~m Telescope}
\label{sec:swope}

As detailed in \citet{krisciunas17}, the CSP-I employed the ``SITe3'' CCD camera and a set 
of Sloan Digital Sky Survey $ugri$ and Johnson $BV$ filters on the LCO Swope telescope to 
obtain optical light curves of the target SNe.  The same setup was used for the first two 
campaigns of the CSP-II.\footnote[29]{All of the $V$-band images taken during the CSP-II were 
obtained with the ``$V$(LC-3009)'' filter \citep[see \S6.1.1 of][for more details]{krisciunas17}.}
However, for the third and fourth campaigns, the SITe3 camera was replaced with a new
camera housing an e2V $4112 \times 4096$ pixel, deep depletion CCD with 15~$\mu$m pixels, four 
read-out amplifiers, and a two-layer anti-etalon, anti-reflection coating.  At the focal plane of the Swope telescope, this corresponds
to a $30\arcmin \times 30\arcmin$ field with 0.\arcsec435 pixels.  This is a larger field than was
required for the CSP-II observations, and so we used only quadrant 3 of the detector since
it had the best linearity characteristics.

\begin{figure}[h]
\epsscale{1.0}
\plotone{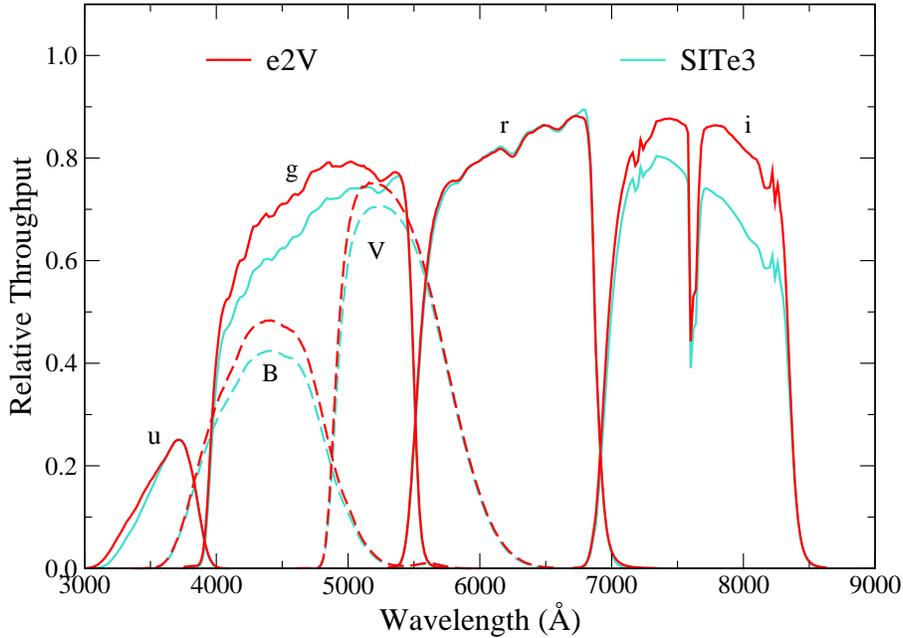} 
\caption{(Cyan) Optical filter response functions for the Swope telescope and SITe3 CCD
camera used during the first two campaigns of the CSP-II. This is the same system 
used during the CSP-I.  (Red) Optical filter response 
functions for the Swope telescope and e2V CCD camera used during the final two 
campaigns of the CSP-II.  The $B$ and $V$ curves are plotted with dashed lines to aid
in their discrimination.  Note that the curves give the total relative 
throughput (telescope~+~filter~+~CCD camera~+~atmosphere) for an airmass of 1.2.}
\label{fig:opt_filters}
\end{figure}

The relative throughputs of the $ugriBV$ passbands with the new e2V CCD were measured
in October 2013 using a monochromator and calibrated detectors \citep{rheault14}.
As expected, the quantum efficiency of the e2V CCD was found to be more uniform and 
sensitive, both in the blue and the red, than the SITe3 CCD.  The response functions
(telescope + filter + CCD camera + atmosphere) for the SITe3 camera are plotted in cyan
in Figure~\ref{fig:opt_filters} .  These are based on measurements made 
at the telescope in January and July 2010 using the same monochromator 
\citep[see][for the calibration details]{stritzinger11b}.  Plotted in red in Figure~\ref{fig:opt_filters} 
are the response functions for the e2V camera.  The curves for both detectors have
been normalized using the $r$ band.  The 
$B$, $g$, $V$, and $i$ filters with the e2V CCD show increased sensitivity 
compared to the $r$ filter than was the case with the SITe3 detector.

The methodology of the CSP photometric reductions is explained in detail in \citet{krisciunas17},
but we briefly reproduce it here for completeness.  First, we solve for color terms and extinction
based on observations on photometric nights of \citet{landolt92} and \citet{smith02} standard stars. 
We then fit the instrumental magnitudes, $ugribv$, via the following equations:

\begin{equation}
u~=~u'~+k_uX~-~\epsilon_u \times (u'~-~g')-~\zeta_u \; ,
\label{eq:u_inst}
\end{equation}
\begin{equation}
g~=~g'~+k_gX~-~\epsilon_g  \times (g'~-~r')-~\zeta_g \; ,
\label{eq:g_inst}
\end{equation}
\begin{equation}
r~=~r'~+k_rX~-~\epsilon_r \times (r'~-~i')-~\zeta_r \; ,
\label{eq:r_inst}
\end{equation}
\begin{equation}
i~=~i'~+k_iX~-~\epsilon_i \times (r'~-~i')-~\zeta_i \; ,
\label{eq:i_inst}
\end{equation}
\begin{equation}
b~=~B~+k_bX~-~\epsilon_b \times (B~-~V)-~\zeta_b \; ,
\label{eq:B_inst}
\end{equation}
\begin{equation}
v~=~V~+k_vX~-~\epsilon_v \times (V~-~i')-~\zeta_v \; ,
\label{eq:V_inst}
\end{equation}

\noindent where $u'g'r'i'BV$ correspond to magnitudes in the standard system, $k_\lambda$ 
are the extinction coefficients, $X$ is the airmass, $\epsilon_\lambda$ are the color terms,
and $\zeta_\lambda$ are the zero points.  During the CSP-I project, $\sim$10 standard stars 
were observed per photometric night to determine the extinction and color terms.  These 
were found to be highly stable over five years \citep[see Figures~4 and~5 of][]{krisciunas17}.  
Hence, a slightly different calibration strategy was implemented for the CSP-II.  The number of 
standard stars observed per night was decreased to $\sim$4--5, but standard-star observations 
were taken on as many clear nights as possible.  As a result, data for determining extinction and 
color terms were obtained on considerably more nights during the CSP-II compared 
to the CSP-I. 

\begin{figure}[h]
\epsscale{0.9}
\plotone{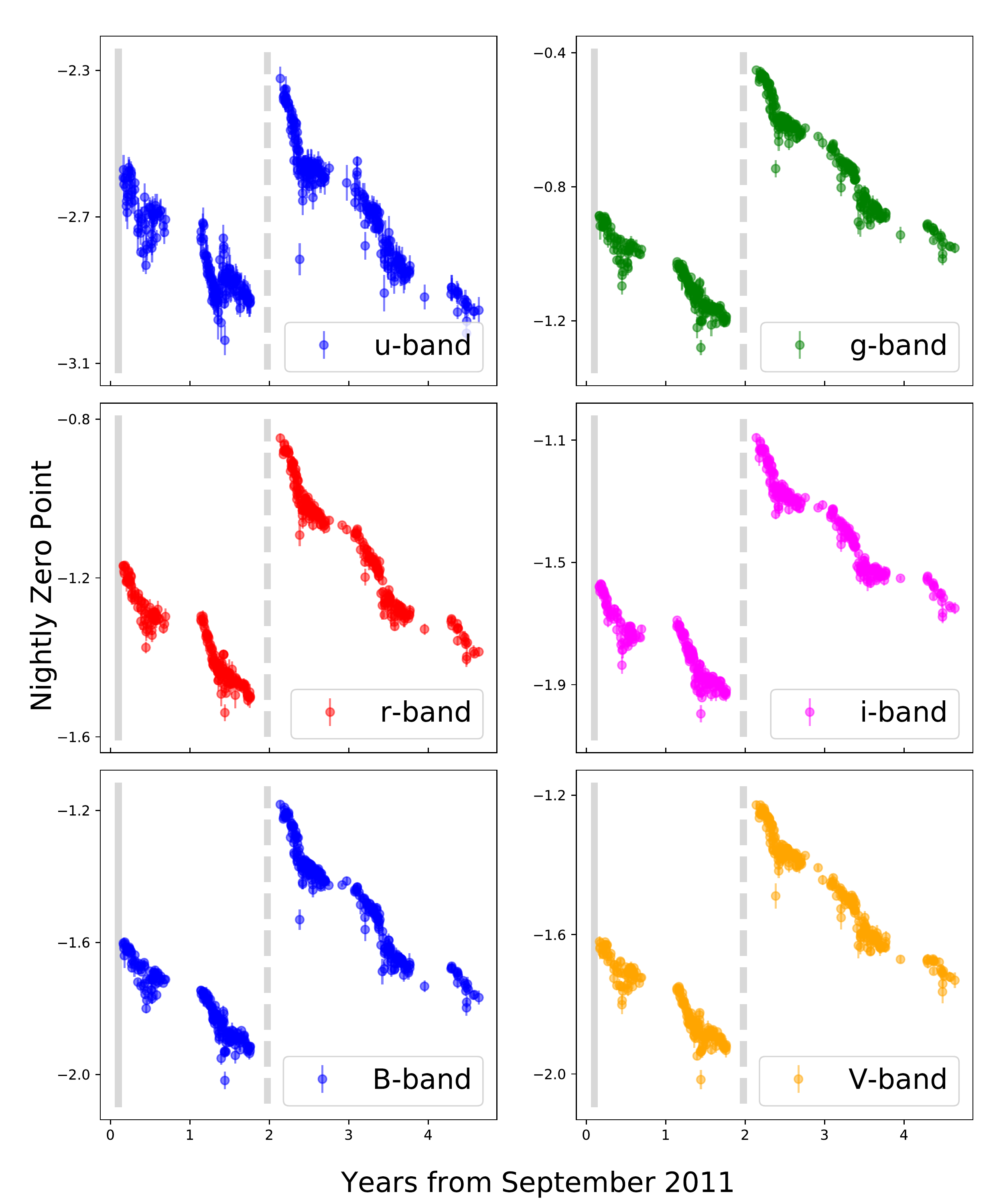} 
\caption{Nightly photometric zero-points derived from observations of local standard stars 
observed on clear nights with the LCO Swope telescope during the four years of the CSP-II.
The solid vertical gray line corresponds to when the primary mirror of the telescope was aluminized;
the dashed vertical gray line indicates when the primary mirror was washed, and also when
the SITe3 detector was replaced by an e2v CCD.  The few points falling below the general trend 
correspond to nights that appeared cloudless but were not photometric.  See text for further
details.}
\label{fig:zpts}
\end{figure}

Figure~\ref{fig:zpts} displays the evolution of the nightly values of the zero points during the CSP-II. 
As mentioned above, the detector in use for the first two years was the same SITe3 CCD used during 
the CSP-I.
The significant jump in sensitivity at the end of year~2 is due to two nearly equal effects.  The first is the 
washing of the primary mirror (indicated by the vertical gray dashed line), and the second is the higher 
quantum efficiency of the e2v CCD, which was put into operation immediately following the mirror 
washing. Each of these effects separately appears to have produced an increase in sensitivity of 
$\sim$50\%. The steady decline in sensitivity observed for both detectors is due to the accumulation 
of dust and aerosols, principally on the primary mirror.  The periodic dips in sensitivity correspond to 
the Chilean midsummer and most likely are due to an increase in atmospheric haze that is typical at 
this time of the year \citep{krisciunas17}.

Table~\ref{tab:ec_ct} compares the final mean extinction and color coefficients for the
Swope~+~e2V system with those of the Swope~+~SITe3.  As was observed for the CSP-I
\citep{krisciunas17}, the extinction terms were remarkably consistent during the four years
of the CSP-II, testifying to the photometric stability of the atmosphere above LCO.  
As expected, some small differences are observed between the color coefficients, with the
largest changes being in the blue (the $u$ and $B$ filters) and the red (the $i$ filter).

\begin{deluxetable} {ccc}
\tabletypesize{\scriptsize}
\tablecolumns{3}
\tablewidth{0pt}
\tablecaption{Optical Photometric Reduction Terms: SITe3 vs. e2V\label{tab:ec_ct}}
\tablehead{
\colhead{Filter} &
\colhead{Swope+SITe3}   &
\colhead{Swope+e2V}
}
\startdata
\multicolumn{3}{c}{Extinction Coefficients\tablenotemark{a}} \\
$u$ & \phn0.511 $\pm$ 0.057 &  0.509 $\pm$ 0.060 \\
$B$ & \phn0.242 $\pm$ 0.022 &  0.233 $\pm$ 0.027 \\
$g$ & \phn0.191 $\pm$ 0.021 & 0.186 $\pm$ 0.027 \\
$V$ & \phn0.144 $\pm$ 0.018 & 0.135 $\pm$ 0.026 \\
$r$ & \phn0.103 $\pm$ 0.019 & 0.094 $\pm$ 0.022 \\
$i$ & \phn0.059 $\pm$ 0.020 & 0.057 $\pm$ 0.020 \\
\hline
\multicolumn{3}{c}{Color Terms\tablenotemark{b}} \\
$u$ & \phn0.046 $\pm$ 0.017 & \phn0.030 $\pm$ 0.020  \\
$B$ & \phn0.061 $\pm$ 0.012 & \phn0.091 $\pm$ 0.015 \\
$g$ & $-$0.014 $\pm$ 0.011 &  $-$0.005 $\pm$ 0.014 \\
$V$ & $-$0.058 $\pm$ 0.011 & $-$0.062 $\pm$ 0.015 \\
$r$ & $-$0.016 $\pm$ 0.015 & $-$0.001 $\pm$ 0.022 \\
$i$ & $-$0.002 $\pm$ 0.015 & \phn0.021 $\pm$ 0.018 \\
\enddata
\tablenotetext{a}{Measured in magnitudes per airmass.  
All unncertainties in this table are the ``standard deviations 
of the distributions,'' not the standard deviations of the means.}
\tablenotetext{b}{See equations~\ref{eq:u_inst}--\ref{eq:V_inst}
for which standard colors are used in combination with these coefficients 
to obtain the color correction terms for the optical photometry.}
\end{deluxetable}

To measure final light curves for the SNe, we first established local sequence stars in 
each of the SN fields from observations of \citet{landolt92} and \citet{smith02} standard stars. 
The underlying host-galaxy light was then subtracted from each SN image using host-galaxy 
reference images acquired after the SN had disappeared.  These images were obtained
mostly with the SITe2 CCD imager on the du~Pont telescope.  However, reference
images for a number of SNe were acquired with the SITe3 CCD on the Swope telescope
in good seeing ($< 1 \arcsec$) conditions.  Magnitudes for the SN were then measured differentially 
with respect to the local sequence stars using point-spread-function (PSF) photometry 
\citep[for further details, see][]{krisciunas17}.  As a matter of policy, the CSP has been publishing 
all of its optical (and NIR) light curves in the {\it natural} photometric system of each 
telescope/instrument/filter combination, which is the ``purest'' form of the data
\citep[see \S5.1 of][]{krisciunas17}.

\subsection{du~Pont 2.5~m Telescope}
\label{sec:dupont_opt}

The LCO du~Pont 2.5~m telescope was used with the facility Tek5 CCD camera to
obtain host-galaxy reference images as well as a small amount of SN follow-up photometry.
As discussed in detail in \S6.1.2 of \citet{krisciunas17}, experiments carried out on two
SN fields observed in both the Swope~~SITe3 and du~Pont~+~Tek5 systems confirm that the 
SN photometry obtained in the $ugriBV$ filters with the du~Pont~+~Tek5 is on substantially the
same natural system as the Swope~+~SITe3 camera.

\section{NIR Photometry}
\label{sec:nir_phot}

\subsection{du~Pont 2.5~m Telescope}
\label{sec:dupont_nir}

NIR imaging in the $YJH$ filters of the CSP-II SNe was obtained with RetroCam on the 
du~Pont telescope.  RetroCam employs a Rockwell $1024 \times 1024$ HAWAII-1 HgCdTe 
array with 18.5~$\mu$m pixels.  This is the same NIR imager used at the 
Swope telescope during the CSP-I.\footnote[30]{Note that all of the RetroCam $J$-band 
observations made during the CSP-II were obtained with the ``$J_{RC2}$'' filter 
\citep[see \S6.2.1 of][for more details]{krisciunas17}.}
RetroCam on the du~Pont telescope gives a field of 
$3.\arcmin4 \times 3.\arcmin4$ and a pixel size of 0.\arcsec201.  For host galaxies that
were larger than the dither pattern, separate sky images were obtained.

\begin{figure}[h]
\epsscale{1.0}
\plotone{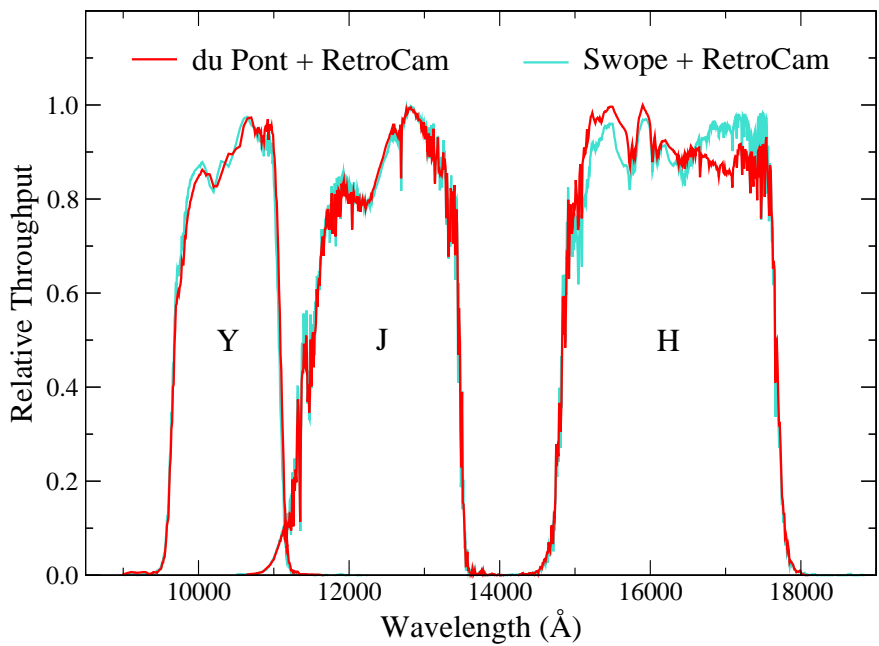} 
\caption{(Cyan) $YJH$ filter response functions for the LCO Swope telescope and RetroCam
imager used during the CSP-I; (Red) the LCO du~Pont 2.5~m telescope and RetroCam used for 
NIR imaging during the CSP-II.  Note that the curves give the total relative throughputs 
(telescope + filter + camera + atmosphere) for an airmass of $\sim$1.2.}
\label{fig:nir_filters}
\end{figure}

In October 2013, \citet{rheault14} made spectrophotometric measurements of the 
RetroCam $YJH$ bandpasses on the du~Pont telescope, which can be compared with 
similar data collected in January 2010 with the same monochrometer when RetroCam 
was on the Swope telescope \citep{krisciunas17}.  This is shown in 
Figure~\ref{fig:nir_filters}, where the total relative throughputs 
(telescope~+~filter~+ camera~+~atmosphere) are shown for RetroCam on the
Swope and du~Pont telescopes.
As may be seen, the transmission functions of the $Y$ and $J$ bands did not change 
significantly between the two measurements.  The most significant difference is in the $H$ filter 
and is likely ascribed to variations in either or both of the reflective and transmissive properties 
of the optics of the two telescopes.  Most importantly, for all three bandpasses, the filter edges 
have not shifted significantly in wavelength. 

The NIR imaging of the SNe was calibrated to local sequence stars
established in the fields of each of the SNe.  Calibration of the local sequence stars 
was carried out through observations on photometric nights of $\sim$4--5 of the 
\citet{persson98} standard stars.
In analogy to the optical filters, the photometric transformation equations for the  
measured instrumental magnitudes, $yjh$, are:

\begin{equation}
y~=~Y_{\rm{std}}~+k_yX~-~\epsilon_y \times (J_{\rm{std}}~-~H_{\rm{std}})-~\zeta_y \; ,
\label{eq:y_inst}
\end{equation}
\begin{equation}
j~=~J_{\rm{std}}~+k_jX~-~\epsilon_j \times (J_{\rm{std}}~-~H_{\rm{std}})-~\zeta_j \; ,
\label{eq:j_inst}
\end{equation}
\begin{equation}
h~=~H_{\rm{std}}~+k_hX~-~\epsilon_h \times (J_{\rm{std}}~-~H_{\rm{std}})-~\zeta_h \; ,
\label{eq:h_inst}
\end{equation}

\noindent where $k_\lambda$ are the extinction coefficients, $X$ is the airmass, 
$\epsilon_\lambda$ are the color terms, and $\zeta_\lambda$ are the zero points.
In these equations, $J_{\rm{std}}$ and $H_{\rm{std}}$ are the magnitudes given in
\citet{persson98}, and $Y_{\rm{std}}$ is the magnitude in the RetroCam $Y$-band standard 
system defined by \citet{krisciunas17}.

\begin{figure}[h]
\epsscale{0.8}
\plotone{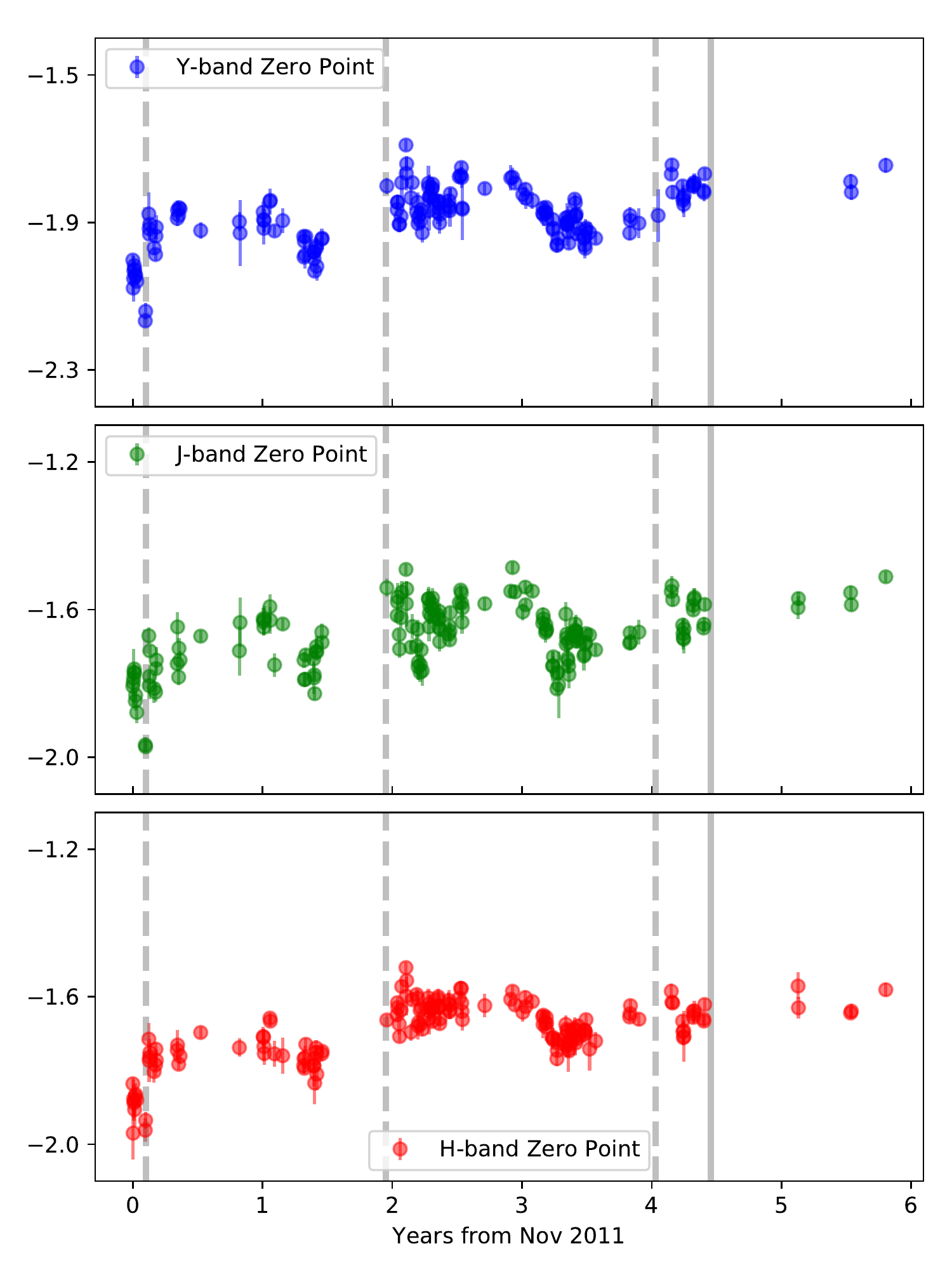} 
\caption{Nightly photometric zero-points derived from observations of standard stars observed 
on clear nights with RetroCam on the LCO 2.5~m du~Pont telescope during the four years of the 
CSP-II. The dashed vertical gray lines indicate when the primary mirror was washed, and the solid 
vertical line corresponds to when the primary mirror was aluminized.}
\label{fig:RC_NIR_ZP}
\end{figure}

Zero points derived from RetroCam during the $4+$ years of the CSP-II are plotted in 
Figure~\ref{fig:RC_NIR_ZP}.  Note that these do not show the steady decline in sensitivity
observed for the CCD detectors on the Swope telescope (cf. Figure~\ref{fig:zpts}).  This is likely 
because the primary mirror of the du~Pont telescope is cleaned with CO$_2$ on 
an approximately weekly basis, whereas the closed-tube design of the Swope telescope does
not permit this.   

Extinction coefficients in the $Y$, $J$, and $H$ filters for RetroCam on the du~Pont telescope
were derived through a simultaneous Markov chain Monte Carlo fitting in the mixture model 
framework \citep{hogg10} of all the CSP-II RetroCam nights.  This procedure assumes a unique 
free parameter for the extinction coefficient, a fixed value of the color term estimated from 
synthetic photometry of model stellar atmospheres (see below), and the nightly zero point 
values.  The resulting values are given in the top half of Table~\ref{tab:nir_ec_ct}.  We note 
that these differ slightly from the extinction coefficients measured with RetroCam on the
Swope telescope during the CSP-I, but fall within the dispersion of the latter values
\citep[see Figure~8 of][]{krisciunas17}. 

\begin{deluxetable} {ccc}
\tabletypesize{\scriptsize}
\tablecolumns{3}
\tablewidth{0pt}
\tablecaption{NIR Photometric Reduction Terms \label{tab:nir_ec_ct}}
\tablehead{
\colhead{Filter} &
\colhead{du~Pont~+~RetroCam}   &
\colhead{Baade~+~FourStar}
}
\startdata
 & \multicolumn{2}{c}{Extinction Coefficients\tablenotemark{a}} \\
$Y$ & \phn0.069 $\pm$ 0.006 &  \ldots \\
$J$ & \phn0.101 $\pm$ 0.006 &  \ldots \\
$H$ & \phn0.056 $\pm$ 0.007 & \ldots \\
\hline
 & \multicolumn{2}{c}{Color Terms\tablenotemark{b}} \\
$Y$ & \phn0.000 & \phn0.106 \\
$J$ & \phn0.019 &  \phn0.001 \\
$H$ & $-$0.039 & $-$0.040 \\
\enddata
\tablenotetext{a}{Measured in magnitudes per airmass.  
Unncertainties in the extinction coefficients are the ``standard deviations 
of the distributions,'' not the standard deviations of the means.}
\tablenotetext{b}{See equations~\ref{eq:y_inst}--\ref{eq:h_inst}
for which standard colors are used in combination with these coefficients 
to obtain the color correction terms for the NIR photometry. The
color terms in this table are estimated from synthetic photometry of \citet{castelli03}
stellar atmosphere models.  The extinction coefficients measured for the
du~Pont~+~RetroCam were assumed for Baade~+~FourStar.}
\end{deluxetable}

The NIR color terms could not be measured at the telescope due to the small color range, 
$+0.19 \leq (J-H) \leq +0.35$~mag, of the \citet{persson98} standard  stars employed for
the nightly photometric calibration.  However, since we have precise spectrophotometric 
measurements of the NIR filter bandpasses, color terms can be estimated from synthetic 
photometry of model stellar atmospheres. This was done for the RetroCam $J_{RC2}$ and
$H$ filters on the du~Pont telescope in Appendix~C of \citet{krisciunas17}, 
and the resulting values, reproduced in Table~\ref{tab:nir_ec_ct}, were shown to be consistent 
with observations made of the red stars listed in Table~3 of \citet{persson98}.  As expected
from the close agreement of the RetroCam $Y$ filter bandpasses on the Swope and 
du~Pont telescopes (see Figure~\ref{fig:nir_filters}), the color term for this filter is negligible 
since the standard system is defined to be the natural system of RetroCam on the Swope 
telescope \citep{krisciunas17}.

Final measurement of the SN light curves was carried out in an analogous way to the
optical photometry.  Local sequence stars were established in each of the SN fields and magnitudes 
for the SNe were measured differentially with respect to these.  

\subsection{Magellan Baade 6.5~m Telescope}
\label{sec:baade_nir}

\begin{figure}[h]
\epsscale{1.0}
\plotone{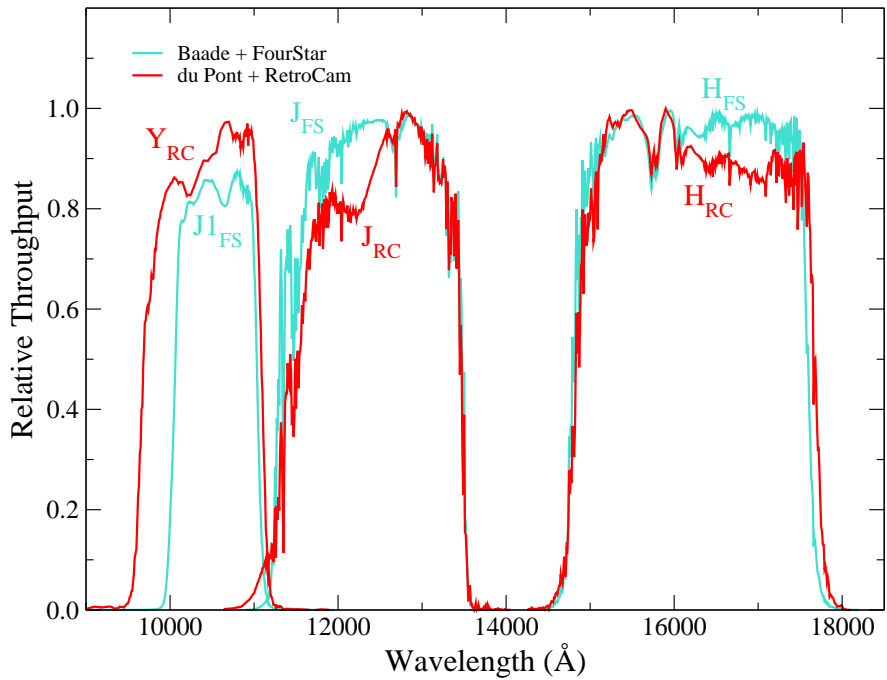} 
\caption{Filter response functions for the Baade telescope and FourStar imager.  
The FourStar filters are plotted in cyan and are compared with the du~Pont + RetroCam 
filters, which are plotted in red.  FourStar does not have a $Y$ filter, but the $J1$ filter 
provides a reasonable approximation \citep[see][]{contreras18}.  Note that the filter curves 
give the total relative throughputs (telescope + filter + camera + atmosphere) for an 
airmass of $\sim$1.2.}
\label{fig:fourstar_filters}
\end{figure}

Host-galaxy reference images were obtained mostly with the FourStar imager \citep{persson13} 
on the Magellan Baade telescope due to the superior throughput and image quality of this 
instrument. (Some reference images were also obtained with RetroCam on the du~Pont 
telescope on nights of excellent seeing.)  As mentioned in \S\ref{sec:nir_imaging}, a few 
active SNe were also observed with FourStar.  FourStar employs four HAWAII-2RG 
detectors that cover a $10.\arcmin8 \times 10.\arcmin8$ field of view with a pixel scale of 0.\arcsec159.  
Filter response functions for the FourStar filters are illustrated in Figure~\ref{fig:fourstar_filters}.  
Note that FourStar does not have a $Y$ filter, but the $J1$ filter provides a useful alternative.
When imaging CSP-II SNe, the target was centered typically in chip~2.  If the host 
galaxy was larger than the dither pattern, the target was alternately positioned in chip~2 
and chip~4 to construct sky images in both chips.

The \citet{persson98} red stars are far too bright to be observed with FourStar on the 
Magellan Baade telescope, and so we must also estimate the color terms for this instrument 
from synthetic photometry of model atmospheres.  The resulting values are listed in 
Table~\ref{tab:nir_ec_ct}. 

\section{First Results}
\label{sec:results}


\begin{figure}[h]
\epsscale{1.1}
\plotone{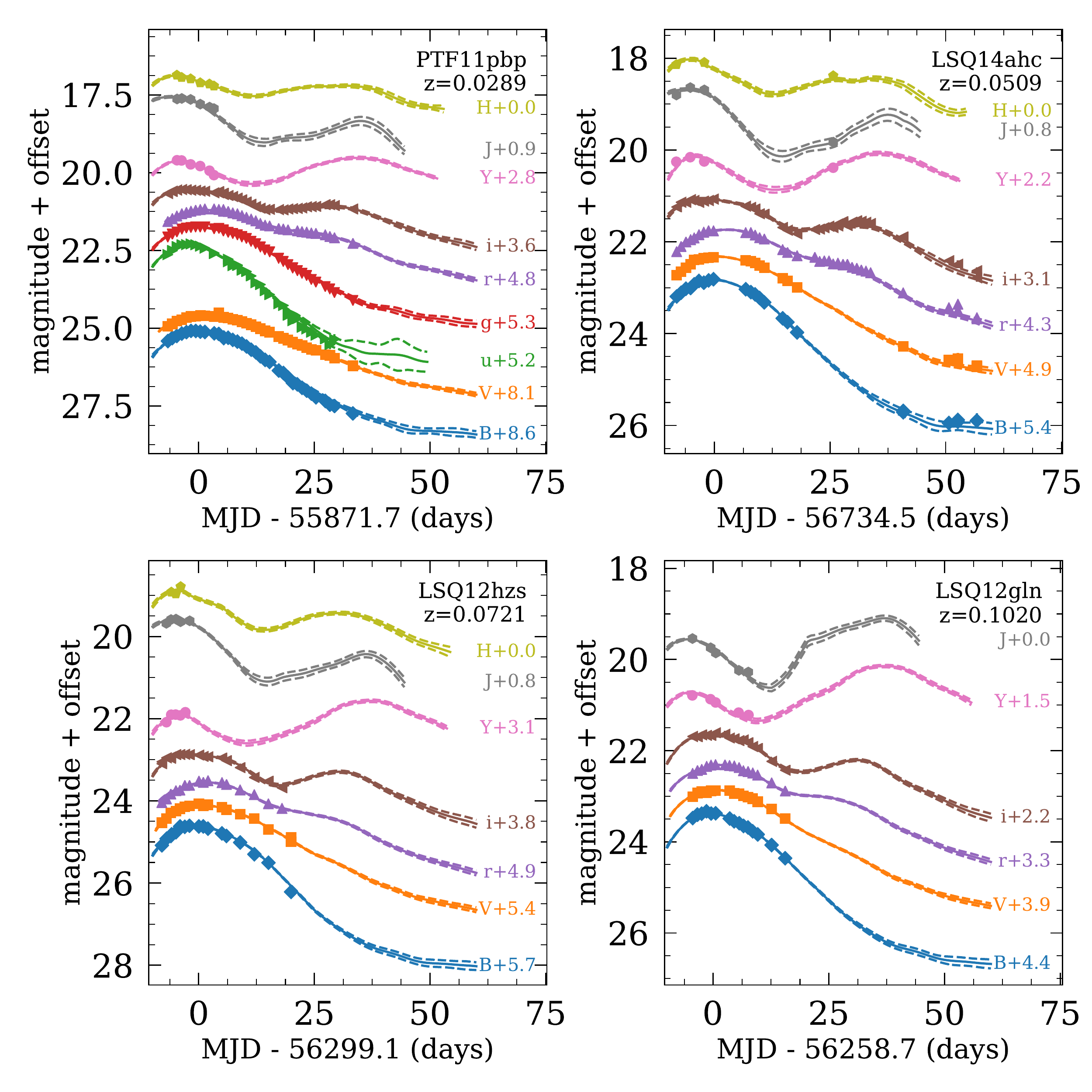} 
\caption{Light curves of four representative SNe~Ia (PTF11pbp, LSQ14ahc, LSQ12hzs, and LSQ12gln) 
are plotted with offsets for clarity. In most cases, the error bars representing the statistical uncertainties 
are smaller than the symbols.  The data have been fitted using the ``max\_model'' option in SNooPy
\citep{burns11} which fits the maximum magnitude in each filter using a template 
generator based on well-observed optical and NIR light curves from the CSP-I
The solid lines represent SNooPy fits to each filter's light curve, while the 
dashed lines represent the $\pm~1\sigma$ errors in the light-curve templates. The filter names and 
magnitude offsets are labeled to the right of each light-curve.}
\label{fig:lcurves}
\end{figure}

\begin{figure}[h]
\epsscale{1.1}
\plotone{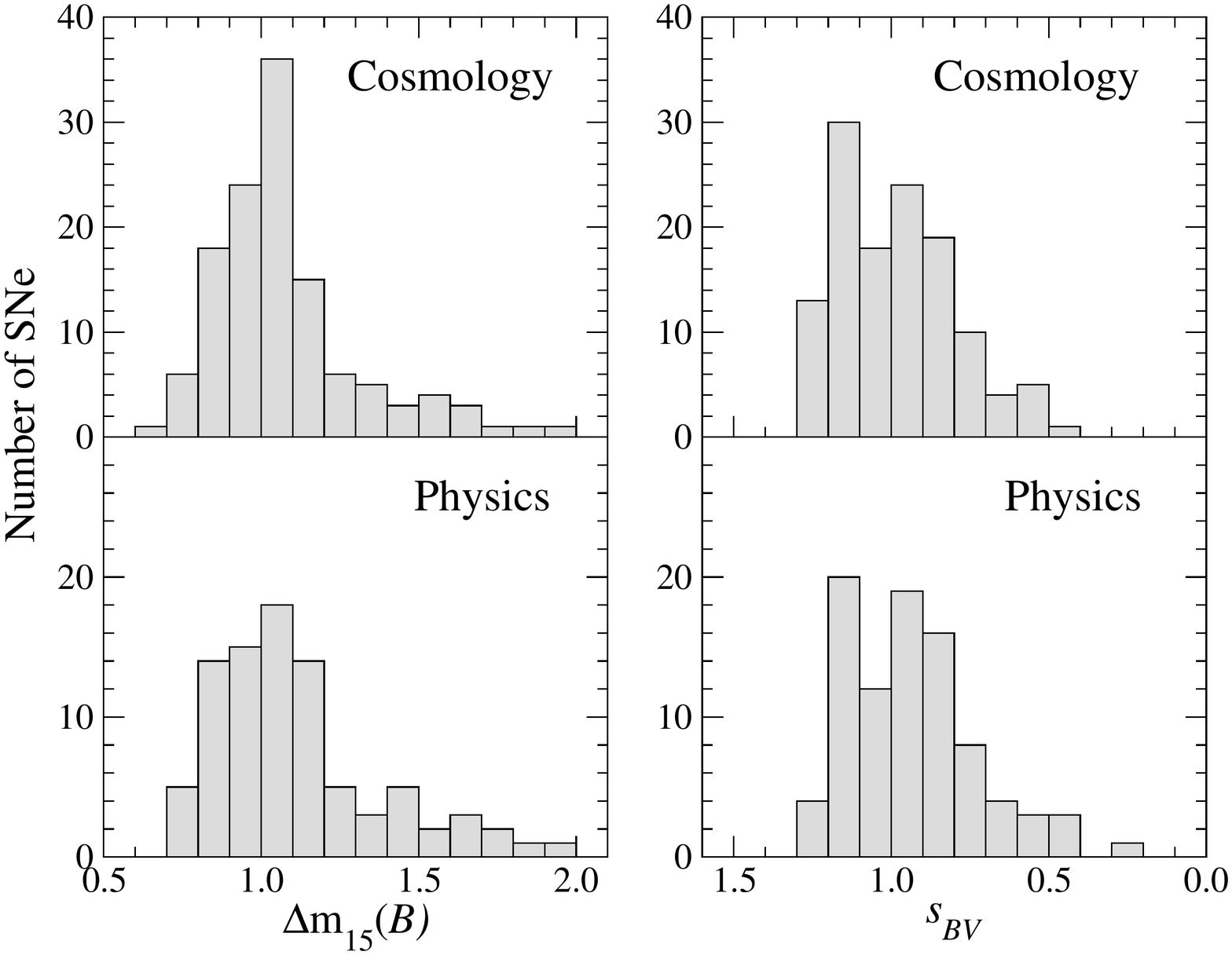} 
\caption{Histograms of the $B$-band decline-rate parameter, $\Delta{\rm m}_{15}(B)$ (left panel), and
the color stretch parameter, $s_{BV}$ (right panel) for the Cosmology and Physics subsamples.}
\label{fig:dm15_sBV}
\end{figure}

Figure~\ref{fig:lcurves} displays optical and NIR light curves of four SNe~Ia in the Cosmology 
subsample observed during the first CSP-II campaign.  These objects were selected as representative 
of both the range of data quality and redshifts covered by the subsample.  The photometric measurements 
have been fitted with the SNooPy \citep[``SuperNovae in object oriented Python'';][]{burns11} package.


In the left half of Figure~\ref{fig:dm15_sBV}, histograms of the $B$-band decline rate, 
$\Delta{\rm m}_{15}(B)$, of the SNe~Ia in the Cosmology and Physics subsamples are plotted as 
determined from the SNooPy template fits.  In the right half of this figure, histograms of the color 
stretch parameter, $s_{BV}$ \citep{burns14}, are also plotted. The distribution of 
the decline-rate parameters for the two subsamples differs only slightly 
in that the Physics subsample has relatively more fast decliners (smaller stretch).  This may be due to that 
fact that 34\% of the SNe Ia in the Physics subsample come from targeted searches monitoring 
predominately massive galaxies, where the vast majority of the fast decliners are found.

The final photometry data release for the full CSP-II sample of 214 SNe~Ia is planned for 2019.
Note that light curves and spectra for a few individual objects observed by the CSP-II have already been
published: LSQ12gdj, a slow-declining, UV-bright SN~Ia \citep{scalzo14}; iPTF13ebh, a transitional 
SN~Ia showing strong NIR \ion{C}{1} lines \citep{hsiao15}; SN~2011iv, a transitional SN~Ia that was
discovered in the same Fornax cluster galaxy that hosted SN~2007on, another transitional SN~Ia observed 
during the CSP-I \citep{gall18}; SN~2012fr, a nearby, peculiar SN~Ia observed within a day of explosion 
\citep{childress13,contreras18}; ASASSN-14lp, a bright SN~Ia also discovered within two days
of outburst \citep{shappee16}; SN~2012Z, a luminous SN~Iax \citep{stritzinger15}; and SN~2013by, 
a Type~IIL SN with a sharp light curve decline after a short, steep plateau/linear decline phase 
\citep{valenti15}.  Optical and NIR photometry of two SNe~Ia observed by the CSP-II, SN~2012ht and 
SN~2015F, that appeared in host galaxies with Cepheid distances is presented in \citet{burns18}.

\section{Conclusions}
\label{sec:conclusions}

This paper has presented a summary of the second phase of the Carnegie Supernova Project,
which was carried out between 2011--2015.  Photometry was obtained for a total
of 214 SNe~Ia with host-galaxy redshifts in the range $0.004 < z < 0.137$.  A ``Cosmology'' subsample 
of 125 SNe with both optical and NIR light curves at a median redshift of $z = 0.056$ 
is described.  These SNe, along with the subsample of 72 SNe~Ia discovered by the La Silla-QUEST
survey, will be used to study the intrinsic precision of SNe~Ia as cosmological distance indicators and
to measure the local value of the Hubble constant.  Light curves and NIR spectroscopy were also 
obtained of a second ``Physics'' 
subsample of 90 SNe~Ia at a median redshift of $z = 0.021$.  This subsample will be used to determine
precise NIR K-corrections and to study the explosion physics and progenitors of SNe~Ia.
The 214 SNe~Ia monitored by the CSP-II combined with the 123 SNe~Ia observed during the 
CSP-I \citep{krisciunas17}  constitutes a sample of more than 300 SNe~Ia with precise light curves
in a well-understood photometric system.  This data set will provide a definitive low-redshift reference 
for future rest frame optical and NIR observations of SNe~Ia at high redshift with next-generation dark 
energy experiments such as the Euclid and WFIRST missions, and the LSST Dark Energy Science 
Collaboration.

\acknowledgments

The work of the CSP-II has been generously supported by the National Science Foundation under 
grants  AST-1008343, AST-1613426, AST-1613455, and AST-1613472.  The CSP-II was also 
supported in part by the Danish Agency for Science and Technology and Innovation through a 
Sapere Aude Level 2 grant. M. Stritzinger acknowledges  funding by a research 
grant (13261) from VILLUM FONDEN.  T. D. is supported by an appointment to the NASA Postdoctoral 
Program at the Goddard Space Flight Center, administered by Universities Space Research Association 
under contract with NASA.



\facilities{Magellan:Baade (IMACS imaging spectrograph, FourStar wide-field near-infrared camera, 
FIRE near-infrared echellette), Magellan:Clay (LDSS3 imaging spectrograph), Swope (SITe3 CCD imager, 
e2v 4K x 4K CCD imager), du~Pont (Tek5 CCD imager, WFCCD imaging spectrograph, RetroCam 
near-infrared imager), Gemini:North (GNIRS near-infrared spectrograph), Gemini:South (FLAMINGOS2),
VLT (ISAAC, MUSE), IRTF (SpeX near-infrared spectrograph), NOT (ALFOSC), Calar Alto 3.5~m (PMAS/PPak),
La Silla-QUEST, CRTS, PTF, iPTF, OGLE, ASAS-SN, PS1, KISS, ISSP, MASTER, SMT)}

\software{SNID \citep{blondin07}, SUPERFIT \citep{howell05}, GELATO \citep{harutyunyan},
SNooPy \citep{burns11}}

\clearpage

\end{document}